\documentclass[useAMS,usenatbib,letterpaper]{mn2e}

\usepackage{times}
\usepackage{psfrag}
\usepackage[dvips]{graphicx}
\usepackage{amssymb}
\usepackage{bm}
\usepackage[french,english]{babel}
\usepackage{color}

\title[Spread spectrum in radio interferometry]
{Spread spectrum for imaging techniques in radio interferometry}

\author[Wiaux et al.]
{Y. Wiaux$^{1,2}$\thanks{E-mail: yves.wiaux@epfl.ch}, G. Puy$^{1}$, Y. Boursier$^{1}$, P. Vandergheynst$^{1}$\\
$^{1}$Institute of Electrical Engineering, Ecole Polytechnique F\'ed\'erale de Lausanne (EPFL), CH-1015 Lausanne, Switzerland\\
$^{2}$Centre for Particle Physics and Phenomenology, Universit\'e catholique de Louvain (UCL), B-1348 Louvain-la-Neuve, Belgium\\}

\begin{document}

\date{\today}

\pagerange{\pageref{firstpage}--\pageref{lastpage}} \pubyear{2009}

\maketitle

\label{firstpage}

\begin{abstract}
We consider the probe of astrophysical signals through radio interferometers with small field of view and baselines with non-negligible and constant component in the pointing direction. In this context, the visibilities measured essentially identify with a noisy and incomplete Fourier coverage of the product of the planar signals with a linear chirp modulation. In light of the recent theory of compressed sensing and in the perspective of defining the best possible imaging techniques for sparse signals, we analyze the related spread spectrum phenomenon and suggest its universality relative to the sparsity dictionary. Our results rely both on theoretical considerations related to the mutual coherence between the sparsity and sensing dictionaries, as well as on numerical simulations.
\end{abstract}


\section{Introduction}

\label{sec:Introduction}Aperture synthesis in radio interferometry is a powerful technique in radio astronomy, allowing observations of the sky with otherwise inaccessible angular resolutions and sensitivities. This technique dates back to more than sixty years ago \citep{ryle46,blythe57,ryle59,ryle60,thompson04}. In this context, the portion of the celestial sphere around the pointing direction tracked by a radio telescope array during observation defines the original real signal or image probed $x$. The field of view observed is limited by a primary beam $A$. Standard interferometers are characterized by a small field of view, so that the signal and the primary beam are assumed to be planar. Considering non-polarized radiation, they respectively read as scalar functions $x(\bm{l})$ and $A(\bm{l})$ of the position $\bm{l}\in\mathbb{R}^{2}$, with components $(l,m)$. Each telescope pair at one instant of observation identifies a baseline defined as the relative position between the two telescopes. To each baseline $\bm{b}_{\lambda}\in\mathbb{R}^{3}$, with components $(u,v,w)$ in units of the signal emission wavelength $\lambda$, is associated one measurement called visibility. In the simplest setting one also considers baselines with negligible component $w$ in the pointing direction of the instrument. Under this additional assumption, if the signal is made up of incoherent sources, each visibility corresponds to the value of the Fourier transform of the signal multiplied by the primary beam at a spatial frequency $\bm{u}\in\mathbb{R}^{2}$, identified by the components $(u,v)$ of the baseline projection on the plane of the signal. Radio-interferometric data are thus identified by incomplete and noisy measurements in the Fourier plane. In the perspective of the reconstruction of the original image, these data define an ill-posed inverse problem.

It is well-known that a large variety of natural signals are sparse or compressible in multi-scale dictionaries, such as wavelet frames. A band-limited signal may be expressed as the $N$-dimensional vector of its values sampled at the Nyquist-Shannon rate. By definition, a signal is sparse in some basis if its expansion contains only a small number $K\ll N$ of non-zero coefficients. More generally it is compressible if its expansion only contains a small number of significant coefficients, i.e. if a large number of its coefficients bear a negligible value. The theory of compressed sensing demonstrates that, for sparse or compressible signals, a small number $M\ll N$ of random measurements, in a sensing basis incoherent with the sparsity basis, will suffice for an accurate and stable reconstruction of the signals \citep{candes06a,candes06b,candes06,donoho06,baraniuk07a,donoho09}. The mutual coherence between two bases may be defined as the maximum complex modulus of the scalar product between unit-norm vectors of the two bases. Random Fourier measurements of a signal sparse in real space are a particular example of a good sensing procedure in this context.

In a very recent work \citep{wiaux09}, we presented results showing that compressed sensing offers powerful image reconstruction techniques for radio-interferometric data, in the case of baselines with negligible component $w$. These techniques are based on global Basis Pursuit (${\rm BP}$) minimization problems, which are solved by convex optimization algorithms. We particularly illustrated the versatility of the scheme relative to the inclusion of specific prior information on the signal in the minimization problems.

In the present work, we raise the important problem of the dependence of the image reconstruction quality as a function of the sparsity basis, or more generally the sparsity dictionary. The larger the typical size of the waveforms constituting the dictionary in which the signal is sparse or compressible in real space, the smaller their extension in the Fourier plane, and the smaller the incoherence between the sparsity and sensing dictionaries. In the context of compressed sensing, a loss of incoherence leads to a degradation of the reconstruction quality for a given sparsity $K$ and a given number $M$ of random measurements.

The detailed structure of radio-interferometric measurements might actually provide a natural response to this issue. The approximation of baselines with negligible component $w$ is a key assumption in order to identify visibilities with Fourier measurements of the original signal. This approximation actually sets a strong constraint on the field of view probed by the interferometer, requiring that it is small enough, not only for the planar approximation of the signal but also to neglect the complete effect of the component $w$ in the visibilities. We relax this approximation and consider radio interferometers with small field of view and baselines with non-negligible component $w$. In this context, each visibility at spatial frequency $\bm{u}$ identifies with the Fourier transform of a complex signal obtained as the product of the original planar signal multiplied by the primary beam with a linear chirp ${\rm e}^{{\rm i}\pi w\vert \bm{l} \vert^2}$ where the norm $\vert \bm{l} \vert$ identifies the distance to the center of the image. Note that this chirp is a priori characterized by a different chirp rate for each baseline: $w=w(\bm{u})$. In the Fourier plane, the modulation amounts to the convolution of the Fourier transform of the chirp with that of the signal multiplied by the primary beam, which inevitably spreads the two-dimensional sample power spectrum of the constitutive waveforms. This spread spectrum phenomenon increases the incoherence between the sparsity and sensing dictionaries.

In this context, we define signals made up of Gaussian waveforms with equal size, identified by a standard deviation $t$, considered as a rough representation of any kind of astrophysical structure. We make the simplifying assumption that all baselines, identified by the spatial frequencies $\bm{u}$, have the same component $w$, so that all visibilities are affected by the same chirp. A theoretical computation of the mutual coherence between the sparsity and sensing dictionaries as a function of $t$ and $w$ naturally proves that the coherence decreases when $w$ increases, through the spread spectrum phenomenon. Our theoretical relation also shows that the coherence may be decreased as close as required to the minimum coherence between the real and Fourier spaces through the use of a large enough chirp rate $w$. In light of the theory of compressed sensing, this suggests some universality of the spread spectrum phenomenon according to which, for a given sparsity $K$ and a given number $M$ of random measurements, the quality of the ${\rm BP}$ reconstruction would not only increase when $w$ increases, but could also be rendered independent of the sparsity dictionary for large enough chirp rate $w$. In other words, the ${\rm BP}$ reconstruction could be made as good for signals sparse in a dictionary of Gaussian waveforms of any size, as for signals sparse in real space. We produce simulations of the signals considered, and perform ${\rm BP}$ reconstructions from noisy visibility measurements. Our results confirm the universality of the spread spectrum phenomenon relative to the sparsity dictionary.

Let us acknowledge the fact that spread spectrum techniques are widely used in telecommunications. In this context, chirp modulations are used in particular to render transmitted signals more robust to noise. In the context of compressed sensing,  spread spectrum techniques using pseudo-random modulations have very recently been highlighted as a means to enhance the signal reconstruction and render it stable relative to noise \citep{naini09}. The use of chirp modulations has also been proposed in compressed sensing radar to produce a convolution of a sparse signal in the sparsity basis itself before performing measurements in that same basis \citep{baraniuk07b, herman09}. In this respect, these techniques are much more related to coded aperture techniques \citep{gottesman89,marcia08} as well as to compressed sensing by random convolution \citep{romberg08} than to spread spectrum techniques.

In Section \ref{sec:Radio interferometry}, we formulate the interferometric inverse problem for image reconstruction on small field of view in the presence of a non-negligible and constant component $w$ of the baselines. In Section \ref{sec:Compressed sensing}, we concisely recall results of the theory of compressed sensing regarding the definition of a sensing basis and the accurate reconstruction of sparse or compressible signals from ${\rm BP}$. In Section \ref{sec:Spread spectrum universality}, we establish the universality of the spread spectrum phenomenon for a sparsity dictionary made up of Gaussian waveforms, both from theoretical considerations and on the basis of simulations. We finally conclude in Section \ref{sec:Conclusion}.

\section{Radio interferometry}
\label{sec:Radio interferometry}

In this section, we recall the general form of the visibility measurements and study the approximation of a small field of view and baselines with non-negligible component $w$. We identify in particular the corresponding spread spectrum phenomenon in the presence of a constant component $w$. We also pose the corresponding interferometric inverse problem for image reconstruction.

\subsection{General visibilities}

\begin{figure}
\begin{center}
\psfrag{o}{ \hspace{-0.1cm}$O$}
\psfrag{north}{ \hspace{0.1cm}north}
\psfrag{e1}{ \hspace{-0.1cm}$\hat{\bm{e}}_1$}
\psfrag{e2}{ \hspace{0cm}$\hat{\bm{e}}_2$}
\psfrag{e3}{ \hspace{-0.4cm}$\hat{\bm{s}}_0,\hat{\bm{e}}_3$}
\psfrag{s}{ \hspace{-0.1cm}{\color{blue}$\hat{\bm{s}}$}}
\psfrag{tau}{ \hspace{0.35cm}{\color{blue}$\bm{\tau}$}}
\psfrag{l}{ \hspace{-0.1cm}{\color{blue}$l$}}
\psfrag{m}{ \hspace{0cm}{\color{blue}$m$}}
\psfrag{n}{ \hspace{0cm}{\color{blue}$n$}}
\psfrag{b}{ \hspace{-0.3cm}{\color{red}$\frac{\bm{b}_{\lambda}}{\vert\bm{b}_{\lambda}\vert}$}}
\psfrag{u}{ \hspace{-0.45cm}{\color{red}$\frac{u}{\vert\bm{b}_{\lambda}\vert}$}}
\psfrag{v}{ \hspace{0.15cm}{\color{red}$\frac{v}{\vert\bm{b}_{\lambda}\vert}$}}
\psfrag{w}{ \hspace{0.15cm}{\color{red}$\frac{w}{\vert\bm{b}_{\lambda}\vert}$}}
\psfrag{x}{ \hspace{-0.3cm}{\color{blue}$x(\bm{\tau})$}}
\psfrag{a}{ \hspace{0cm}{\color{red}$A(\bm{\tau})$}}
\includegraphics[trim = 0cm 4cm 0cm 0cm, clip, width=6cm,keepaspectratio]{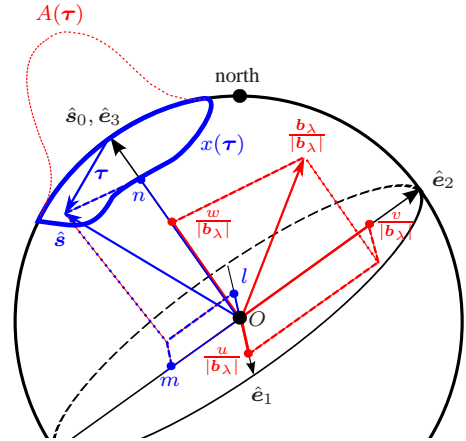}
\end{center}
\caption{\label{fig:notations}Illustration of notations. The signal probed $x(\bm{\tau})$, with $\bm{\tau}\equiv\hat{\bm{s}}-\hat{\bm{s}}_0$, extends in any direction on the celestial sphere ${\rm S}^{2}$ identified by unit vectors $\hat{\bm{s}}\in\mathbb{R}^{3}$, around the pointing direction $\hat{\bm{s}}_0\in\mathbb{R}^{3}$. The field of view of the radio interferometer is set by the primary beam $A(\bm{\tau})$. The baselines also characterizing the interferometer are vectors $\bm{b}_{\lambda}\in\mathbb{R}^{3}$ with norm $\vert\bm{b}_{\lambda}\vert$. The coordinate system ${\rm R}:(O,\hat{\bm{e}}_1, \hat{\bm{e}}_2, \hat{\bm{e}}_3)$ in $\mathbb{R}^{3}$ is illustrated, with $\hat{\bm{e}}_3$ identified with $\hat{\bm{s}}_0$, $\hat{\bm{e}}_2$ pointing toward north, and $\hat{\bm{e}}_1$ pointing toward east. The components of the direction vectors $\hat{\bm{s}}$ on the sphere are denoted by $(l,m,n)$, while those of a baseline are denoted by $(u,v,w)$.}
\end{figure}
In a tracking configuration, all radio telescopes of an interferometric array point in the same direction on the unit celestial sphere ${\rm S}^{2}$  identified by a unit vector $\hat{\bm{s}}_0\in\mathbb{R}^{3}$. Arbitrary directions are denoted by unit vectors $\hat{\bm{s}}\in\mathbb{R}^{3}$. The field of view observed is limited by a so-called primary beam $A(\bm{\tau})$, where the vector $\bm{\tau}\equiv\hat{\bm{s}}-\hat{\bm{s}}_0\in\mathbb{R}^{3}$ identifies directions relative to the pointing direction (see Fig. \ref{fig:notations}). The size of its angular support is essentially inversely proportional to the size of the dishes of the telescopes. We consider a monochromatic signal $x(\bm{\tau})$ with an emission wavelength $\lambda$, and made up of incoherent sources (see Fig. \ref{fig:notations}). Also considering non-polarized radiation, both the signal and the primary beam are scalar square-integrable functions on the sphere, $x(\bm{\tau}),A(\bm{\tau})\in {\rm L}^2({\rm S}^{2},{\rm d}\Omega(\bm{\tau}))$, where ${\rm d}\Omega(\bm{\tau})$ stands for the invariant measure on the sphere. At each instant of observation, each telescope pair measures a complex visibility defined as the correlation between incoming electric fields at the positions of the two telescopes. This visibility only depends on the relative position between the two telescopes, defined as a baseline. The baseline in units of $\lambda$ is denoted by a vector $\bm{b}_{\lambda}\in\mathbb{R}^{3}$ (see Fig. \ref{fig:notations}). Each visibility measured reads in general form as \citep{thompson04}:
\begin{equation}
y\left(\bm{b}_{\lambda}\right) \equiv \int_{{\rm S}^{2}} Ax\left(\bm{\tau}\right) {\rm e}^{-2{\rm i}\pi \bm{b}_{\lambda} \bm{\cdot} \bm{\tau}} \: {\rm d}\Omega\left(\bm{\tau} \right). \label{ri1}
\end{equation}
We consider the conventional right-handed Cartesian coordinate system ${\rm R}:(O,\hat{\bm{e}}_1, \hat{\bm{e}}_2, \hat{\bm{e}}_3)$ in $\mathbb{R}^{3}$ centered on the Earth. The direction $\hat{\bm{e}}_3$ identifies with the pointing direction $\hat{\bm{s}}_0$. The directions $\hat{\bm{e}}_1$ and $\hat{\bm{e}}_2$ identify the orthogonal plane, parallel to the plane tangent to the celestial sphere at the pointing direction, with $\hat{\bm{e}}_2$ pointing toward north, and $\hat{\bm{e}}_1$ pointing toward east. The unit vectors $\hat{\bm{s}}$ may be identified by their components $(l,m,n)$ in ${\rm R}$ (see Fig. \ref{fig:notations}). Each point is thus obviously identified by the two-dimensional vector $\bm{l}\in\mathbb{R}^{2}$ with components $(l,m)$ in the plane $(O,\hat{\bm{e}}_1, \hat{\bm{e}}_2)$, while the component in direction $\hat{\bm{e}}_3$ reads as $n({\bm{l}})=(1-\vert\bm{l}\vert^2)^{1/2}$ with the norm $\vert\bm{l}\vert\equiv(l^2+m^2)^{1/2}$ identifying the distance to the origin in the plane $(O,\hat{\bm{e}}_1, \hat{\bm{e}}_2)$. In particular, $\hat{\bm{s}}_0$ bears by definition the components $(0,0,1)$ so that $\bm{\tau}$ reads as $(\bm{l},n({\bm{l}})-1)$. Signal and primary beam may also be seen as functions of this two-dimensional vector, respectively $x(\bm{l})$ and $A(\bm{l})$. The invariant measure on the sphere reads as ${\rm d}\Omega(\bm{\tau})\equiv n^{-1}(\bm{l}){\rm d}^2\bm{l}$, where  ${\rm d}^2\bm{l}\equiv{\rm d}l{\rm d}m$ stands for the canonical invariant measure in the plane $(O,\hat{\bm{e}}_1, \hat{\bm{e}}_2)$. The components of $\bm{b}_{\lambda}$ are standardly denoted by $(u,v,w)$, so that its norm reads as $\vert\bm{b}_{\lambda}\vert\equiv(u^2+v^2+w^2)^{1/2}$ (see Fig. \ref{fig:notations}). These may also be divided into the two-dimensional vector $\bm{u}\in\mathbb{R}^{2}$ with component $(u,v)$ in the plane $(O,\hat{\bm{e}}_1, \hat{\bm{e}}_2)$ and the component $w$ in direction $\hat{\bm{e}}_3$, i.e. in the pointing direction of the radio interferometer.

The general  form (\ref{ri1}) of the visibilities reads in terms of these coordinates as:
\begin{equation}
y\left(\bm{u},w\right) = \int_{{\rm D}^{2}} Ax\left(\bm{l}\right) {\rm e}^{-2{\rm i}\pi \left[\bm{u} \bm{\cdot} \bm{l} + w \left(n\left(\bm{l}\right)-1\right) \right]} \: n^{-1}\left(\bm{l}\right){\rm d}^2\bm{l}, \label{ri1'}
\end{equation}
where the integration is performed inside the unit disc ${\rm D}^{2}$ on the plane $(O,\hat{\bm{e}}_1, \hat{\bm{e}}_2)$.

In the course of an observation, for each telescope pair, the baseline components $u$, $v$, and $w$ all follow a sinusoidal dependence in time thanks to the Earth rotation, with specific parameters linked to the parameters of observation. The total number of points $\left(\bm{u},w\right)$ probed by all telescope pairs of the array during the observation provides some coverage in $\mathbb{R}^{3}$ characterizing the interferometer.

\subsection{Small field of view and non-negligible $w$}

Note that for a given telescope array, the baseline component $w$ may be seen as a function $w=w({\bm{u}})$ of the two-dimensional vector $\bm{u}$ with an implicit dependence on the distance between the two telescopes considered. Each visibility may thus be seen as a function of the two-dimensional vectors $\bm{u}$: $y(\bm{u},w)=y(\bm{u})$. From this point of view, the points $\bm{u}$ probed by all telescope pairs during the observation provide some coverage of the plane $(O,\hat{\bm{e}}_1, \hat{\bm{e}}_2)$ also characterizing the interferometer. The two-dimensional vector $\bm{u}$ associated with each telescope pair actually runs over an arc of ellipse in this plane.

In this context, we make the two assumptions of small field of view and baselines with non-negligible component $w$. Firstly, we consider a standard interferometer with a primary beam of small enough angular support, so that the field of view may be identified with a patch of the tangent plane at $\hat{\bm{s}}_0$. Technically this amounts to assume that $\vert \bm{l} \vert^2\ll1$ on the field of view so that the zero order expansion $n(\bm{l})\simeq1$ is valid in the invariant measure on the sphere, which therefore identifies with that on the plane: ${\rm d}\Omega(\bm{\tau})={\rm d}^2\bm{l}$. The signal and the primary beam hence become scalar square-integrable functions in the plane $x(\bm{l}),A(\bm{l})\in {\rm L}^2(\mathbb{R}^{2},{\rm d}^2\bm{l})$. Secondly, we consider interferometers with large enough values of the  $w(\bm{u})$, so that the same approximation on $n(\bm{l})$ does not hold for the phase of the imaginary exponential \citep{thompson04}. In this case any term discarded in $n(\bm{l})$ should be small relative to $w^{-1}(\bm{u})$. We assume however that the further constraint on the field of view $\vert \bm{l} \vert^4\ll w^{-1}(\bm{u})$ holds for all $\bm{u}$, so that the second order approximation $n(\bm{l})\simeq1-\vert \bm{l} \vert^2/2$ is valid in the imaginary exponential. Under these conditions, each visibility takes the form:
\begin{equation}
y\left(\bm{u}\right) = \int_{{\rm D}^{2}} Ax\left(\bm{l}\right) {\rm e}^{{\rm i}\pi w(\bm{u})\vert\bm{l}\vert^2} {\rm e}^{-2{\rm i}\pi \bm{u}\bm{\cdot}\bm{l}} \: {\rm d}^{2}\bm{l}.\label{ri2}
\end{equation}
In other words the visibility in $\bm{u}$ identifies with the value in $\bm{u}$ of the two-dimensional Fourier transform of a complex signal obtained as the product of the original planar signal $Ax(\bm{l})$ with a linear chirp modulation $C^{(w(\bm{u}))}(\vert\bm{l}\vert)\equiv {\rm e}^{{\rm i}\pi w(\bm{u})\vert\bm{l}\vert^2}$. This chirp is characterized by a chirp rate $w(\bm{u})$ depending on $\bm{u}$, and $\vert\bm{l}\vert$ now identifies the distance to the origin in the plane of the signal, i.e. to the point identified by $\hat{\bm{s}}_0$.

Note that in the further approximation that $w(\bm{u})=0$  for all $\bm{u}$, the chirp indeed disappears and relation (\ref{ri3}) reduces to the standard van Cittert-Zernike theorem. This theorem states that the visibilities measured identify with the two-dimensional Fourier transform of the signal multiplied by the primary beam: $y(\bm{u}) = \widehat{Ax}(\bm{u})$  \citep{wiaux09}.

The fact that baselines may have non-negligible component $w$ is at the origin of important challenges from the computational point of view. This issue is related to the fact that the explicit computation of simulated visibilities in (\ref{ri2}) from a signal and the corresponding inverse problem have a large complexity. In the case of baselines with negligible component $w(\bm{u})$ at each $\bm{u}$, the simple two-dimensional Fourier transform relation between signal and visibilities allows the use of the Fast Fourier Transform (FFT), which substantially lightens the computation. This is not possible anymore in the case of baselines with non-negligible component $w(\bm{u})$ as the chirp modulation is characterized by a chirp rate explicitly dependent on $\bm{u}$, and therefore needs to be imposed separately for each visibility. A number of algorithms have been studied in this case, among which faceting algorithms as well as the very recent $w$-projection algorithm \citep{cornwell08a}. In one word this $w$-projection algorithm simply computes a unique FFT, after which the visibility at each $\bm{u}$ is obtained by the evaluation, at this $\bm{u}$, of the convolution of the Fourier transform of the signal with that of the chirp with chirp rate  $w(\bm{u})$. Considerations on the extension of the corresponding convolution kernel in the Fourier plane allows to drastically reduce the computational load in practice. These algorithmic issues are however not our concern here. We only aim at discussing how this natural and compulsory chirp modulation may drastically enhance the quality of signal reconstruction. 

\subsection{Constant $w$ and spread spectrum phenomenon}

\label{sub:Constant $w$ and spread spectrum phenomenon}We already acknowledged that, in the course of an observation, the two-dimensional vector $\bm{u}$ associated with each telescope pair runs over an arc of ellipse, and the corresponding component $w$ follows a sinusoidal evolution. However, the whole baseline distribution in $\mathbb{R}^{3}$ is extremely dependent on the specific configuration of the radio telescope array under consideration. Visibilities from various interferometers may also be combined, as well as visibilities from the same interferometer with different pointing directions through the mosaicking technique \citep{thompson04}. From this point of view the baseline distributions are rather flexible.

Relying on this flexibility we assume that all baselines have the same component in the pointing direction: $w(\bm{u})=w$.\footnote{Note for illustration that, for a unique interferometer pointing to the north celestial pole, each telescope pair would exhibit constant $w$ during observation. For particular configurations of telescopes in two east-west linear arrays, all telescope pairs with each telescope belonging to a different array would produce baselines with identical value of $w$.} This assumption of constant component $w$ allows us to discard considerations related to specific interferometers. Moreover, it also allows us to study the impact of a chirp modulation on the basis of a large range of simulations at light computational load. Under this assumption all visibilities indeed identify with the values of the two-dimensional Fourier transform of a complex signal obtained as the product of the original planar signal $Ax(\bm{l})$ with the same linear chirp $C^{(w)}(\vert\bm{l}\vert) = {\rm e}^{{\rm i}\pi w\vert\bm{l}\vert^2}$ characterized by the chirp rate $w$ independent of $\bm{u}$:
\begin{equation}
y\left(\bm{u}\right) = \widehat{C^{(w)}Ax}\left(\bm{u}\right).\label{ri3}
\end{equation}

Note that the multiplication by the chirp modulation amounts to the convolution of the Fourier transform of the chirp with that of the original signal multiplied by the primary beam: $\widehat{C^{(w)}Ax}=\widehat{C^{(w)}}\star\widehat{Ax}$. This convolution inevitably spreads the two-dimensional sample power spectrum of the signal multiplied by the primary beam, i.e. the square modulus of its Fourier transform, while preserving its norm.

In particular, the original signal $x(\bm{l})$ and the primary beam $A(\bm{l})$ can be approximated by band-limited functions on the finite field of view set by the primary beam itself. A linear chirp $C^{(w)}(\vert\bm{l}\vert)$ with chirp rate $w$ is characterized by an instantaneous frequency $w\bm{l}$ at position $\bm{l}$. On the finite field of view set by the primary beam it is therefore approximately a band-limited function.  In each basis direction, the band limit of the signal after convolution $B^{(C^{(w)}Ax)}$ is the sum of the individual band limits of the original signal multiplied by the primary beam $B^{(Ax)}$ and of the chirp modulation $B^{(C^{(w)})}$: $B^{(C^{(w)}Ax)}=B^{(C^{(w)})}+B^{(Ax)}$. This simple consideration already quantifies the spread spectrum phenomenon associated with the chirp modulation. Note for completeness that the primary beam is by definition limited at much lower frequencies than the signal so that it does not significantly affect its band limit: $B^{(Ax)}=B^{(A)}+B^{(x)}\simeq B^{(x)}$.

\subsection{Interferometric inverse problem}

\label{sub:Interferometric inverse problem}The band-limited functions considered are completely identified by their Nyquist-Shannon sampling on a discrete uniform grid of $N=N^{1/2}\times N^{1/2}$ points $\bm{l}_{i}\in\mathbb{R}^{2}$ in real space with $1\leq i\leq N$. The sampled signal is denoted by a vector $\bm{x}\in\mathbb{R}^{N}\equiv\{x_{i}\equiv x(\bm{l}_{i})\}_{1\leq i\leq N}$ and the primary beam is denoted by the vector $\bm{A}\in\mathbb{R}^{N}\equiv\{A_{i}\equiv A(\bm{l}_{i})\}_{1\leq i\leq N}$. The chirp is complex and reads as the vector $\bm{C}^{(w)}\in\mathbb{C}^{N}\equiv\{C^{(w)}_{i}\equiv C^{(w)}(\vert \bm{l}_{i} \vert)\}_{1\leq i\leq N}$. Because of the assumed finite field of view, the functions may equivalently be described by their complex Fourier coefficients on a discrete uniform grid of $N=N^{1/2}\times N^{1/2}$ spatial frequencies $\bm{u}_{i}$ with $1\leq i\leq N$. This grid is limited at some maximum frequency defining the band limit. Due to the fact that the chirp is complex, the Fourier transform of the product $C^{(w)}Ax$ does not bear any specific symmetry property in the Fourier plane.

As in \citet{wiaux09}, we assume that the spatial frequencies $\bm{u}$ probed by all telescope pairs during the observation belong to the discrete grid of points $\bm{u}_{i}$. The Fourier coverage provided by the $M/2$ spatial frequencies probed $\bm{u}_b$, with $1\leq b\leq M/2$, can simply be identified by a binary mask in the Fourier plane equal to $1$ for each spatial frequency probed and $0$ otherwise. The visibilities measured may be denoted by a vector of $M/2$ complex Fourier coefficients $\bm{y}\in\mathbb{C}^{M/2}\equiv\{y_{b}\equiv y(\bm{u}_b)\}_{1\leq b\leq M/2}$, possibly affected by complex noise of astrophysical or instrumental origin, identified by the vector $\bm{n}\in\mathbb{C}^{M/2}\equiv\{n_{b}\equiv n(\bm{u}_{b})\}_{1\leq b\leq M/2}$. Formally, the measured visibilities may equivalently be denoted by a vector of $M$ real measures $\bm{y}\in\mathbb{R}^{M}\equiv\{y_{r}\}_{1\leq r\leq M}$ consisting of the real and imaginary parts of the complex measures, affected by the corresponding real noise values $\bm{n}\in\mathbb{R}^{M}\equiv\{n_{r}\}_{1\leq r\leq M}$.

In this discrete setting, the Fourier coverage is in general incomplete in the sense that the number of real constraints $M$ is smaller than the number of unknowns $N$: $M<N$. An ill-posed inverse problem is thus defined for the reconstruction of the signal $\bm{x}$ from the measured visibilities $\bm{y}$:
\begin{equation}
\bm{y}\equiv\mathsf{\Phi}^{(w)}\bm{x}+\bm{n} \textnormal{ with } \mathsf{\Phi}^{(w)}\equiv\mathsf{MFC}^{(w)}\mathsf{A},\label{ri4}
\end{equation}
where the matrix $\mathsf{\Phi}^{(w)}\in\mathbb{C}^{(M/2)\times N}$ identifies the complete linear relation between the signal and the visibilities. The matrix $\mathsf{A}\in\mathbb{R}^{N\times N}\equiv\{A_{ij}\equiv A_{i}\delta_{ij}\}_{1\leq i,j\leq N}$ is the diagonal matrix implementing the primary beam. The matrix $\mathsf{C}^{(w)}\in\mathbb{C}^{N\times N}\equiv\{C^{(w)}_{ij}\equiv C^{(w)}_{i}\delta_{ij}\}_{1\leq i,j\leq N}$ is the diagonal matrix implementing the chirp modulation. The unitary matrix $\mathsf{F}\in\mathbb{C}^{N\times N}\equiv\{F_{ij}\equiv {\rm e}^{-2{\rm i}\pi\bm{u}_i\bm{\cdot}\bm{x}_j}/N^{1/2}\}_{1\leq i,j\leq N}$ implements the discrete Fourier transform providing the Fourier coefficients. The matrix $\mathsf{M}\in\mathbb{R}^{(M/2)\times N}\equiv\{M_{bj}\}_{1\leq b\leq M/2;1\leq j\leq N}$ is the rectangular binary matrix implementing the mask characterizing the interferometer. It contains only one non-zero value on each line, at the index of the Fourier coefficient corresponding to each of the spatial frequencies probed $\bm{u}$.

In the perspective of the reconstruction of the signal $\bm{x}$, relation (\ref{ri4}) represents the measurement constraint. We take a statistical point of view and consider independent Gaussian noise on each real measure $y_{r}$. Considering a candidate reconstruction $\bar{\bm{x}}$, the residual noise reads as $\bar{\bm{n}}^{(w)}\equiv\bm{y}-\mathsf{\Phi}^{(w)}\bar{\bm{x}}$. The residual noise level estimator, defined as twice the negative logarithm of the likelihood associated with $\bar{\bm{x}}$, reads as
\begin{equation}
\chi^{2}\left(\bar{\bm{x}};\mathsf{\Phi}^{(w)},\bm{y}\right)\equiv\sum_{r=1}^{M}\left(\frac{\bar{n}^{(w)}_{r}}{\sigma^{(n_r)}}\right)^{2},\label{ri5}
\end{equation}
with $\bar{\bm{n}}^{(w)}\in\mathbb{R}^{M}\equiv\{\bar{n}^{(w)}_{r}\}_{1\leq r\leq M}$, and where $\sigma^{(n_r)}$ stands for the standard deviation of the noise component $n_r$. This noise level estimator follows a chi-square distribution with $M$ degrees of freedom. Typically, this estimator should be minimized by the good candidate reconstruction. The measurement constraint on the reconstruction may be defined as a bound $\chi^{2}\left(\bar{\bm{x}};\mathsf{\Phi}^{(w)},\bm{y}\right)\leq\epsilon^{2}$, with $\epsilon^{2}$ corresponding to some percentile of the chi-square distribution. Let us note that the expectation value of the $\chi^{2}$ is equal to its number of degrees of freedom $M$, while its standard deviation is $(2M)^{1/2}$. In other words, for a large number of degrees of freedom the distribution is extremely peaked around its expectation value. This fact is related to the well-known phenomenon of the concentration of measure \citep{candes06b}. As a consequence, the value $\epsilon^{2}$ is anyway extremely close to $M$.

In this context, many signals may formally satisfy the measurement constraint. A regularization scheme that encompasses enough prior information on the original signal is needed in order to find a unique solution. All image reconstruction algorithms will differ through the kind of regularization considered.

\section{Compressed sensing}
\label{sec:Compressed sensing}

In this section, we concisely recall the inverse problem for sparse signals considered in the compressed sensing framework. We describe the ${\rm BP}$ minimization problem set for reconstruction and recall the randomness and incoherence properties for a suitable sensing basis.
\subsection{Inverse problem for sparse signals}

In the framework of compressed sensing \citep{candes06a,candes06b,candes06,donoho06,baraniuk07a,donoho09} the signals probed are firstly assumed to be sparse or compressible in some basis. Technically, one considers a real signal identified by its Nyquist-Shannon sampling, denoted by the vector $\bm{x}\in\mathbb{R}^{N}\equiv\{x_{i}\}_{1\leq i\leq N}$. A real orthonormal basis $\mathsf{\Psi}\in\mathbb{R}^{N\times N}\equiv\{\Psi_{ij}\}_{1\leq i,j\leq N}$ is also considered, in which the decomposition $\bm{\alpha}\in\mathbb{R}^{N}\equiv\{\alpha_i\}_{1\leq i\leq N}$ of the signal is defined by
\begin{equation}
\bm{x} \equiv \mathsf{\Psi}\bm{\alpha}.\label{cs1}
\end{equation}
The signal is said to be sparse or compressible in this basis if it only contains a small number $K \ll N$ of non-zero or significant coefficients $\alpha_i$ respectively.

Secondly, the signal is assumed to be probed by $M$ real linear measurements denoted by a vector $\bm{y}\in\mathbb{R}^{M}\equiv\{y_{r}\}_{1\leq r\leq M}$ in some real sensing basis $\mathsf{\Phi}\in\mathbb{R}^{M\times N}\equiv\{\Phi_{rj}\}_{1\leq r\leq M;1\leq j\leq N}$ and possibly affected by independent and identically distributed noise $\bm{n}\in\mathbb{R}^{M}\equiv\{n_{r}\}_{1\leq r\leq M}$. This defines an inverse problem
\begin{equation}
\bm{y}\equiv\mathsf{\Theta}\bm{\alpha}+\bm{n}\textnormal{ with }\mathsf{\Theta}\equiv\mathsf{\Phi\Psi}\in\mathbb{R}^{M\times N},\label{cs2}
\end{equation}
where the matrix $\mathsf{\Theta}$ identifies the sensing basis as seen from the sparsity itself. The number $M$ of constraints is typically assumed to be smaller than the dimension $N$ of the vector defining the signal, so that the inverse problem is ill-posed.

\subsection{${\rm BP}$ reconstruction}

The framework proposes the global ${\rm BP}$ minimization problem for the signal recovery. This problem regularizes the originally ill-posed inverse problem by an explicit sparsity or compressibility prior on the signal. In the presence of noise, the so-called Basis Pursuit denoise (${\rm BP}_{\epsilon}$) problem is the minimization of the $\ell_{1}$ norm $\vert\vert\bar{\bm{\alpha}}\vert\vert_{1}$ of the coefficients of the signal in the sparsity basis under a constraint on the $\ell_{2}$ norm $\vert\vert\bar{\bm{n}}\vert\vert_{2}$ of the residual noise, with $\bar{\bm{n}}\equiv \bm{y}-\mathsf{\Theta}\bar{\bm{\alpha}}$:
\begin{equation}
\min_{\bar{\bm{\alpha}}\in\mathbb{R}^{N}}\vert\vert\bar{\bm{\alpha}}\vert\vert_{1}\textnormal{ subject to }\vert\vert \bm{y}-\mathsf{\Theta}\bar{\bm{\alpha}}\vert\vert_{2}\leq\epsilon.\label{cs3}
\end{equation}
Let us recall that the $\ell_{1}$ norm of $\bar{\bm{\alpha}}$ is defined as $\vert\vert \bar{\bm{\alpha}}\vert\vert_{1}\equiv\sum_{i=1}^{N}\vert \bar{\alpha}_{i}\vert$. The  $\ell_{2}$ norm of the residual noise is the standard norm of the corresponding vector: $\vert\vert \bar{\bm{n}} \vert\vert_{2}\equiv(\sum_{i=1}^{M}\vert \bar{n}_{i}\vert^{2})^{1/2}$. In these relations, the notation $\vert a \vert$ for a scalar  $a$ stands for the complex modulus when applied to a complex number and the absolute value when applied to a real number. The ${\rm BP}_{\epsilon}$ problem is solved by application of non-linear and iterative convex optimization algorithms \citep{combettes07,vandenBerg08}.

Note that the $\ell_{2}$ norm term in the ${\rm BP}_{\epsilon}$ problem is identical to a bound on the $\chi^{2}$ distribution with $M$ degrees of freedom governing the noise level estimator.

\subsection{Randomness and incoherence}

Among other approaches \citep{donoho09}, the theory of compressed sensing defines the explicit restricted isometry property (RIP) that the matrix $\mathsf{\Theta}=\mathsf{\Phi\Psi}$ should satisfy in order to allow an accurate recovery of sparse or compressible signals \citep{candes06a,candes06b,candes06}. In that regard, the issue of the design of the sensing matrix $\mathsf{\Phi}$ is of course fundamental. The theory offers various ways to design suitable sensing matrices, showing in particular that a small number of measurements is required relative to a naive Nyquist-Shannon sampling: $M\ll N$.

One can actually show that incoherence of $\mathsf{\Phi}$ with the sparsity or compressibility basis $\mathsf{\Psi}$ and randomness of the measurements will ensure that the RIP is satisfied with overwhelming probability, provided that the number of measurements is large enough relative to the sparsity $K$ considered \citep{candes06b,candes06,candes08}. In particular, measurements may be performed through a uniform random selection of Fourier frequencies. In this case, the precise condition for the RIP depends on the degree of incoherence between the Fourier basis and the sparsity or compressibility basis. The mutual coherence $\mu$ between the orthonormal Fourier and sparsity bases may be defined as the maximum complex modulus of the scalar product between basis vectors of the two bases. If the unit-normed basis vectors corresponding to the lines of $\mathsf{F}$ and the columns of $\mathsf{\Psi}$ are denoted by $\{\bm{f}_{i}\}_{1\leq i\leq N}$ and $\{\bm{\psi}_{j}\}_{1\leq j\leq N}$, the mutual coherence between the bases reads as:
\begin{equation}
\mu\left(\mathsf{F},\mathsf{\Psi}\right)\equiv\max_{1\leq i,j\leq N}\vert \bm{f}_{i} \bm{\cdot} \bm{\psi}_{j} \vert. \label{cs4}
\end{equation}
In other words this mutual coherence identifies with the maximum complex modulus of the Fourier coefficient values of the sparsity basis vectors. Note that the inverse of the mutual coherence, $\mu^{-1}(\mathsf{F},\mathsf{\Psi})$, can also be called the mutual incoherence between the Fourier and sparsity bases.

In this context, the RIP is already satisfied for a small number of measurements satisfying the constraint
\begin{equation}
K\leq\frac{cM}{N\mu^{2}\left(\mathsf{F},\mathsf{\Psi}\right)\ln^{4}N},\label{cs5}
\end{equation}
for some constant $c$. Under this condition,\footnote{Let us also acknowledge the fact that this bound is not tight. Empirical results \citep{candes05,lustig07} suggest that ratios $M/K$ between $3$ and $5$ already ensure the expected reconstruction quality.} the ${\rm BP}_{\epsilon}$ reconstruction of signals well approximated by a $K$ sparse signal is shown to be accurate and stable relative to noise as well as relative to compressibility, in the sense of a departure from exact sparsity. Let us emphasize the fact that the mutual coherence plays an essential role in relation (\ref{cs5}) as for fixed $M$ the sparsity recovered increases with the mutual incoherence, as $K\propto\mu^{-2}(\mathsf{F},\mathsf{\Psi})$.

The incoherence is notably maximum between the Fourier basis and the real space basis identified by a sparsity matrix $\mathsf{\Psi}\equiv\mathsf{\Delta}$ made up of unit spikes: $\mu(\mathsf{F},\mathsf{\Delta})=N^{-1/2}$. In the continuous limit $N\rightarrow\infty$ the real space basis identifies with a Dirac basis and the maximum incoherence is infinite, corresponding to zero coherence:
\begin{equation}
\lim_{N\rightarrow\infty}\mu(\mathsf{F},\mathsf{\Delta})=0.\label{cs6}
\end{equation}
Hence, for a fixed number $M$ of pure Fourier measurements, the sparsity recovered $K$ is maximum for signals sparse in real space.

\section{Spread spectrum universality}
\label{sec:Spread spectrum universality}

In this section, we define simple astrophysical signals sparse in a dictionary of Gaussian waveforms, and explicitly identify our sensing dictionary for radio-interferometric measurements. We compute the theoretical coherence between the sparsity and sensing dictionaries as a function of the chirp rate and of the size of the Gaussian waveforms, which suggests the universality of the spread spectrum phenomenon relative to the sparsity dictionary.  We define an observational set up, and describe our simulations and specific reconstruction procedures. We finally expose the results of the analysis, and comment on the need for future work along these lines.

\subsection{Sparsity and sensing dictionaries}

We consider simple astrophysical signals sparse in a dictionary of Gaussian waveforms, all with equal and fixed size given by a standard deviation $t\in\mathbb{R}_+$. These signals are assumed to be probed by radio interferometers with a Gaussian primary beam $A^{(t_0)}$ with a size identified by a standard deviation $t_0\in\mathbb{R}_+$. The corresponding matrix thus reads as $\mathsf{A}^{(t_0)}$. The value $t_0$ sets the size the field of view of interest, which must naturally be larger than the size of the Gaussian waveforms: $t_0>t$.  We consider a small field of view and baselines with a non-negligible constant component $w$, so that the visibilities measured  $y_b=y(\bm{u}_b)$ take the form (\ref{ri3}). Relying on the previous discussion relative to the flexibility of realistic baseline distributions, we also assume that the spatial frequencies $\bm{u}_b$ probed arise from a uniform random selection of Fourier frequencies. As for the assumption of constant $w$, this allows us to discard considerations related to specific interferometers. It also allows us to place our discussion in a setting which complies directly with the requirement of the theory of compressed sensing for random measurements. Let us however recall that specific deterministic distributions of a low number of linear measurements might in fact also allow accurate signal reconstruction in the context of compressed sensing \citep{matei08}.

In this context, the interferometric inverse problem (\ref{ri4}) simply arises from a uniform random selection of spatial frequencies with a sensing dictionary $\mathsf{\Phi}\equiv \mathsf{\Phi}^{(w,t_0)}\equiv\mathsf{MFC}^{(w)}\mathsf{A}^{(t_0)}$ also parametrized by the size $t_0$ of the primary beam. The sparsity dictionary identifying Gaussian waveforms of size $t$ may be denoted by $\mathsf{\Psi}\equiv \mathsf{\Psi}^{(t)}\equiv \mathsf{\Gamma}^{(t)}$ . The sensing dictionary as seen from the sparsity dictionary itself therefore reads as $\mathsf{\Theta}^{(w,t_0,t)}\equiv\mathsf{\Phi}^{(w,t_0)}\mathsf{\Gamma}^{(t)}\equiv\mathsf{MFC}^{(w)}\mathsf{A}^{(t_0)}\mathsf{\Gamma}^{(t)}$.

Let us emphasize the fact that the sparsity dictionary is obviously not an orthogonal basis. Moreover, the sensing dictionary does not correspond anymore exactly to a random selection of vectors in an orthogonal basis. No precise compressed sensing result similar to the bound (\ref{cs5}) relating sparsity and mutual coherence between such dictionaries was yet obtained. However, in the line of first results already proved with redundant dictionaries \citep{rauhut08}, one can intuitively conjecture that a bound similar to (\ref{cs5}) holds in our context, still exhibiting the same trade-off between sparsity and mutual coherence. In the perspective of assessing the ${\rm BP}_{\epsilon}$ reconstruction quality, it is therefore essential to understand how the mutual coherence between the sensing and sparsity dictionaries depends on the size $t$ of the Gaussian waveforms and on the chirp rate $w$.

\subsection{Theoretical coherence}

\label{sub:Theoretical coherence}After normalization of the vectors of the sparsity and sensing dictionaries in $\ell_2$ norm, a simple analytical computation gives the mutual coherence between the sensing and sparsity dictionaries as:
\begin{equation}
\mu\left(\mathsf{FC}^{(w)}\mathsf{A}^{(t_0)},\mathsf{\Gamma}^{(t)}\right)= \frac{2tt_0}{t^2+t_0^2} \left[ 1+\left( \frac{2\pi w t^2 t_0^2}{t^2 + t_0^2} \right)^2 \right]^{-\frac{1}{2}}.\label{ssu1}\end{equation}
Note that one can formally re-organise the decomposition of the matrix $\mathsf{\Theta}^{(w,t_0,t)}$ into modified sensing and sparsity dictionaries, respectively $\tilde{\mathsf{\Phi}}\equiv\mathsf{MF}$ and $\tilde{\mathsf{\Psi}}^{(w,t_0,t)}\equiv\mathsf{C}^{(w)}\mathsf{A}^{(t_0)}\mathsf{\Gamma}^{(t)}$. In this perspective, the mutual coherence $\mu(\mathsf{FC}^{(w)}\mathsf{A}^{(t_0)},\mathsf{\Gamma}^{(t)})$ identifies with the maximum complex modulus of the Fourier coefficient values of the modified sparsity dictionary vectors, which depend on the chirp modulation and on the primary beam. Consequently, the requirement for a lower coherence can be fulfilled by an operation that spreads the two-dimensional sample power spectrum of the signal in the Fourier plane, while preserving its norm.

The relation (\ref{ssu1}) is valid both for finite $N$ as well as in the continuous limit $N\rightarrow\infty$. In this continuous limit, analyzing the evolution of the coherence as a function of the parameters $t$, $t_0$, and $w$ is very enlightening.

Firstly, we consider finite non-zero values of the size $t_0$ of the primary beam and of the chirp rate $w$. When the size $t$ of the Gaussian waveforms tends to zero, the mutual coherence tends to zero:
\begin{equation}
\lim_{t\rightarrow 0}\mu\left(\mathsf{FC}^{(w)}\mathsf{A}^{(t_0)},\mathsf{\Gamma}^{(t)}\right)=0  \textnormal{ for all }w,t_0\in\mathbb{R}_+.\label{ssu2}
\end{equation}
In the absence of chirp modulation and primary beam, this null value is expected from relation (\ref{cs6}) as the modified sparsity dictionary identifies with the Dirac basis. But we also show that the mutual coherence gets to this minimum value independently of the chirp rate $w$ in this limit, and less importantly independently of the size of the primary beam $t_0$.

Secondly, we consider finite non-zero values of the size $t_0$ of the primary beam and of the size $t$ of the Gaussian waveforms, with $t<t_0$. In the absence of chirp, i.e. for $w=0$, the coherence is strictly positive: $\mu(\mathsf{FC}^{(0)}\mathsf{A}^{(t_0)},\mathsf{\Gamma}^{(t)})= 2tt_0/(t^2+t_0^2)$. The finite size $t$ of a Gaussian waveform implies a Gaussian profile of its Fourier transform with a standard deviation $(2\pi t)^{-1}$. The larger $t$, the smaller the extent of the two-dimensional sample power spectrum in the Fourier plane, and the larger the coherence. As discussed in Section \ref{sub:Constant $w$ and spread spectrum phenomenon}, in the presence of a chirp, i.e. for $w\neq0$, the spectrum is spread by means of the corresponding convolution in the Fourier plane, while the norm of the signal multiplied by the primary beam is preserved. This spread spectrum phenomenon reduces the coherence which, in light of relation (\ref{cs5}), should enhance the quality of the ${\rm BP}_{\epsilon}$ reconstruction in the context of compressed sensing. In the limit of an infinite chirp rate, the coherence also tends to its minimum null value independently of the size $t$ of the Gaussian waveforms:
\begin{equation}
\lim_{w\rightarrow \infty}\mu\left(\mathsf{FC}^{(w)}\mathsf{A}^{(t_0)},\mathsf{\Gamma}^{(t)}\right)=0 \textnormal{ for all }t,t_0\in\mathbb{R}_+.\label{ssu3}
\end{equation}
This limit provides a strong result in the sense that the incoherence lost by considering a sparsity dictionary of Gaussian waveforms of arbitrary non-zero size $t$ may be completely recovered thanks to a chirp modulation with high enough chirp rate $w$. Still from relation (\ref{cs5}), this result suggests the universality of the spread spectrum phenomenon according to which the quality of the ${\rm BP}_{\epsilon}$ reconstruction can be rendered independent of the sparsity dictionary for a large enough component $w$ of the baselines.

\subsection{Observational set up}

\begin{figure*}
\begin{center}
\includegraphics[width=4cm,keepaspectratio]{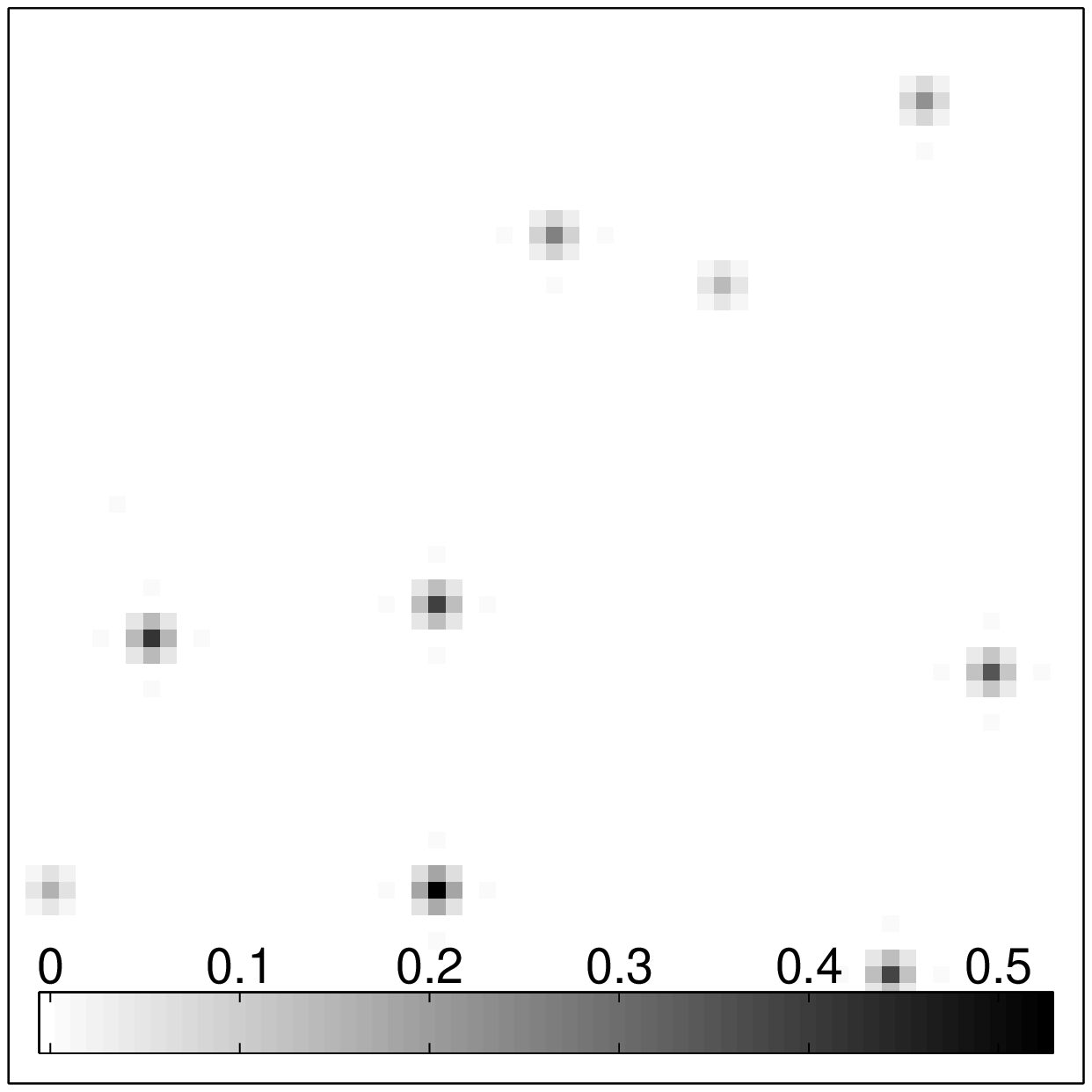}
\hspace{0.2cm}
\includegraphics[width=4cm,keepaspectratio]{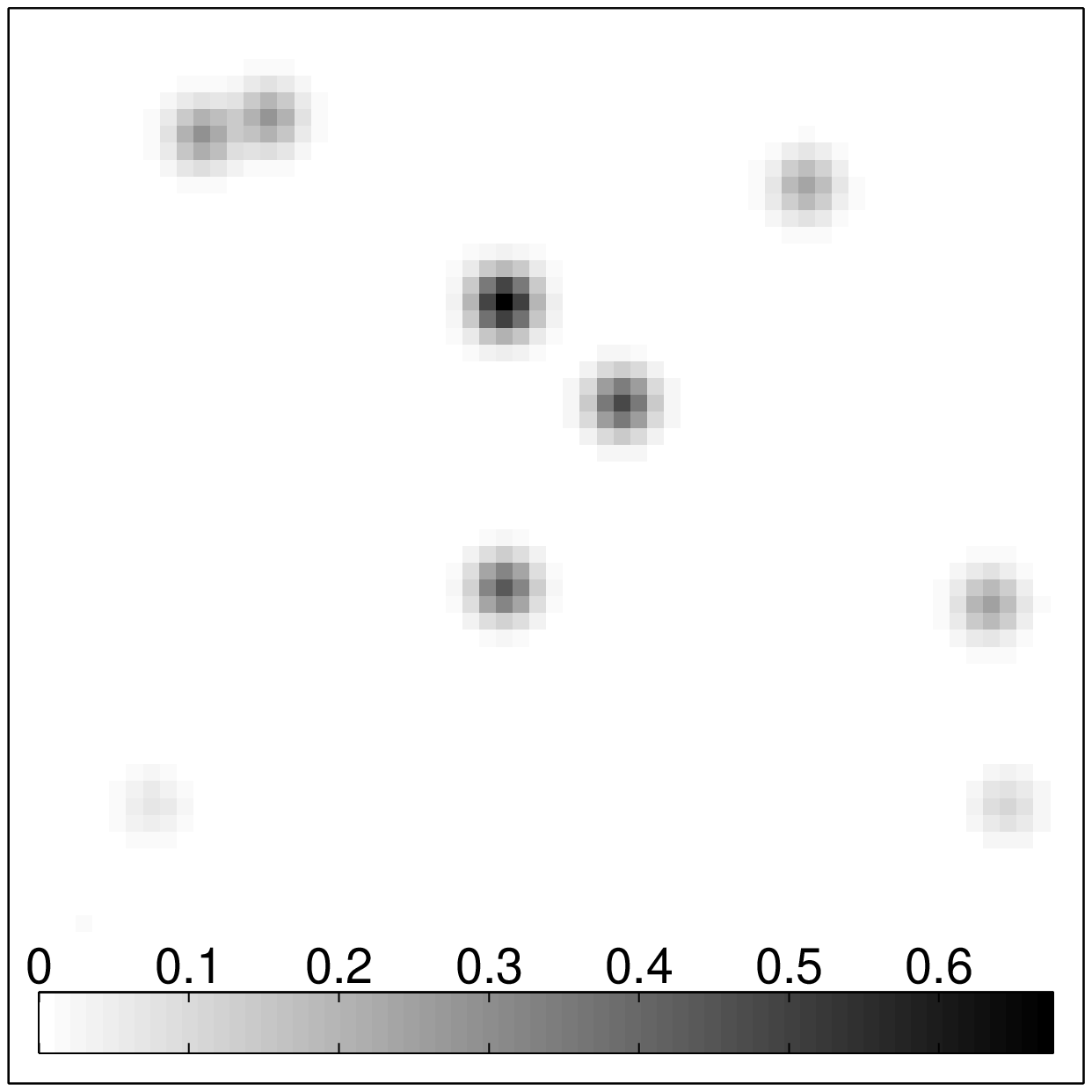}
\hspace{0.2cm}
\includegraphics[width=4cm,keepaspectratio]{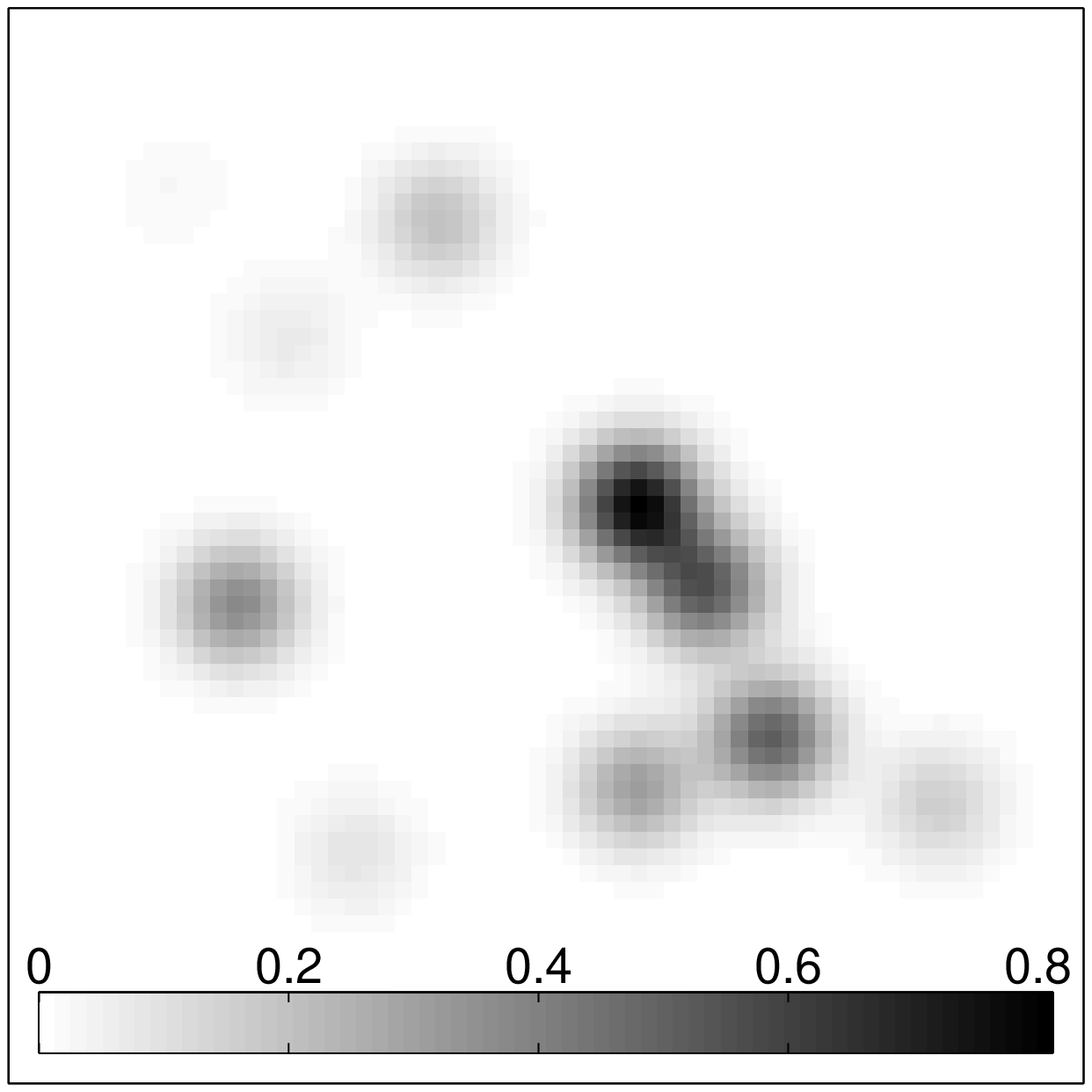}
\hspace{0.2cm}
\includegraphics[width=4cm,keepaspectratio]{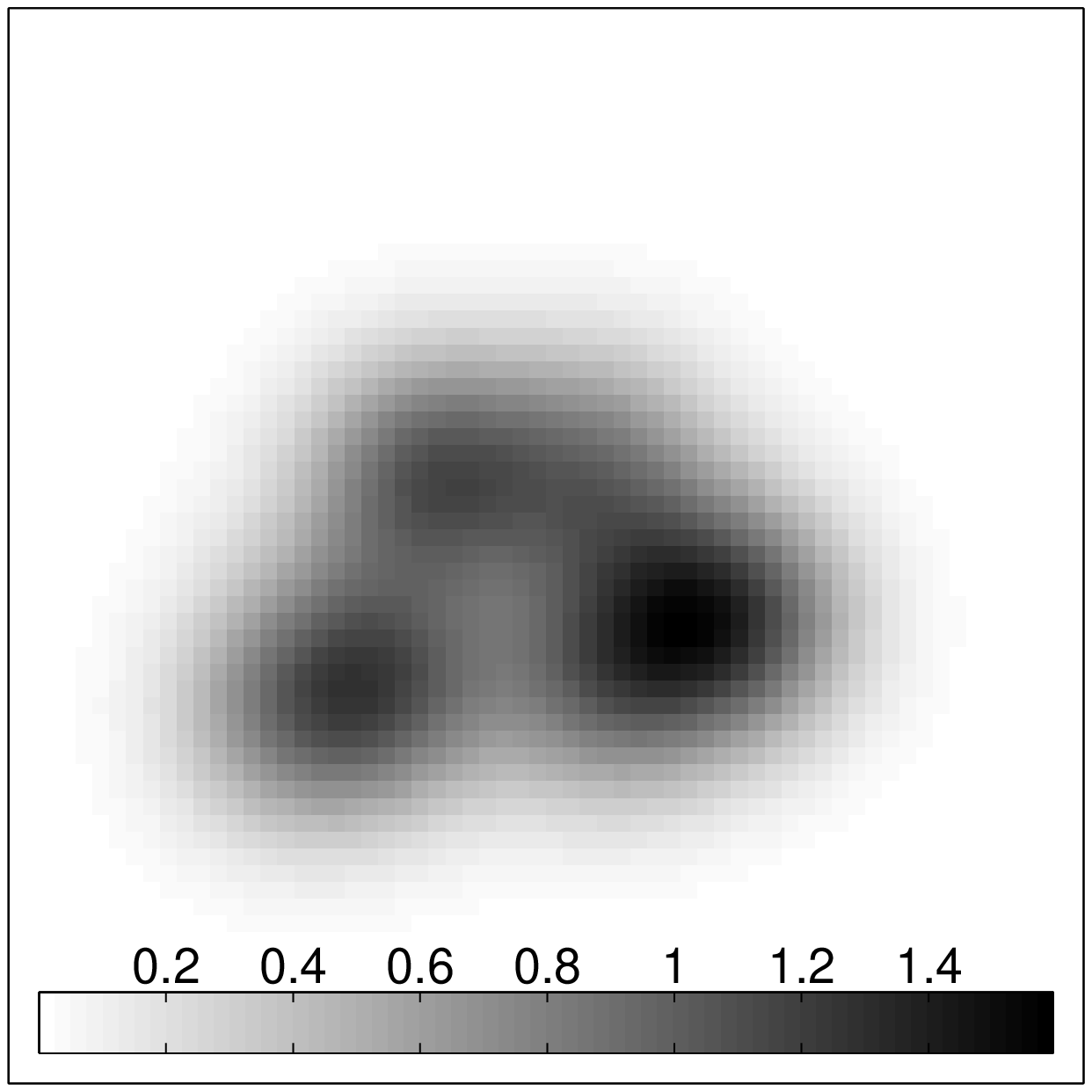}\\
\vspace{0.2cm}
\psfrag{Frequency}{ \hspace{-0.4cm}{\scriptsize Spatial frequency}}
\psfrag{Power spectrum}{\hspace{-0.3cm}{\scriptsize Power spectrum (${\rm iu}^2$)}}
\psfrag{Chirp}{ \hspace{0cm}{\scriptsize \color{blue}Chirp}}
\psfrag{Spread}{\hspace{2cm}{\scriptsize  \color{red}Spread}}
\psfrag{Original}{\hspace{1cm}{\scriptsize  \color{black}Original}}
\includegraphics[width=4cm,keepaspectratio]{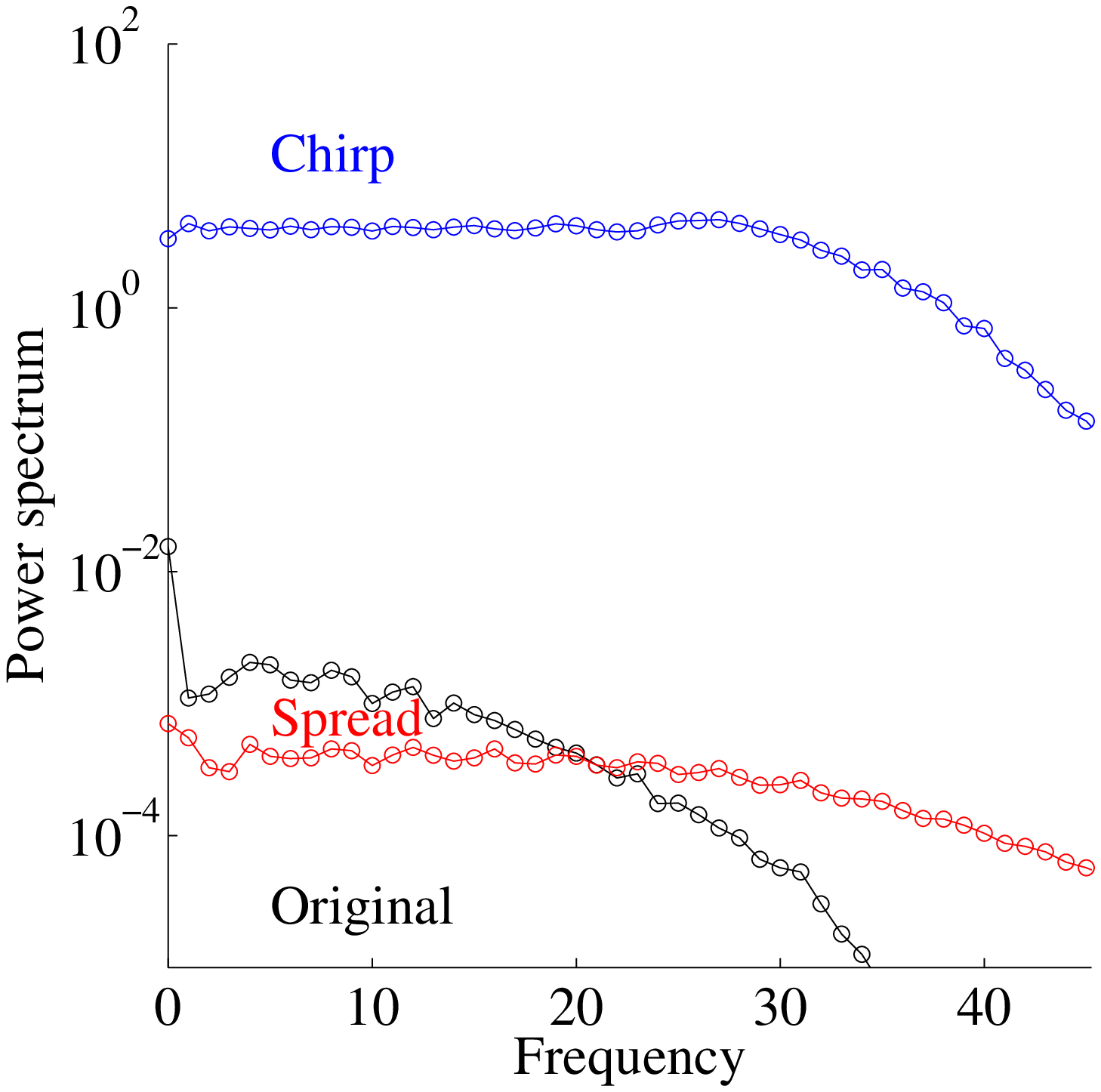}
\hspace{0.2cm}
\psfrag{Frequency}{ \hspace{-0.4cm}{\scriptsize Spatial frequency}}
\psfrag{Power spectrum}{\hspace{-0.3cm}{\scriptsize Power spectrum (${\rm iu}^2$)}}
\psfrag{Chirp}{ \hspace{0cm}{\scriptsize \color{blue}Chirp}}
\psfrag{Spread}{\hspace{2.1cm}{\scriptsize \color{red}Spread}}
\psfrag{Original}{\hspace{1.3cm}{\scriptsize Original}}
\includegraphics[width=4cm,keepaspectratio]{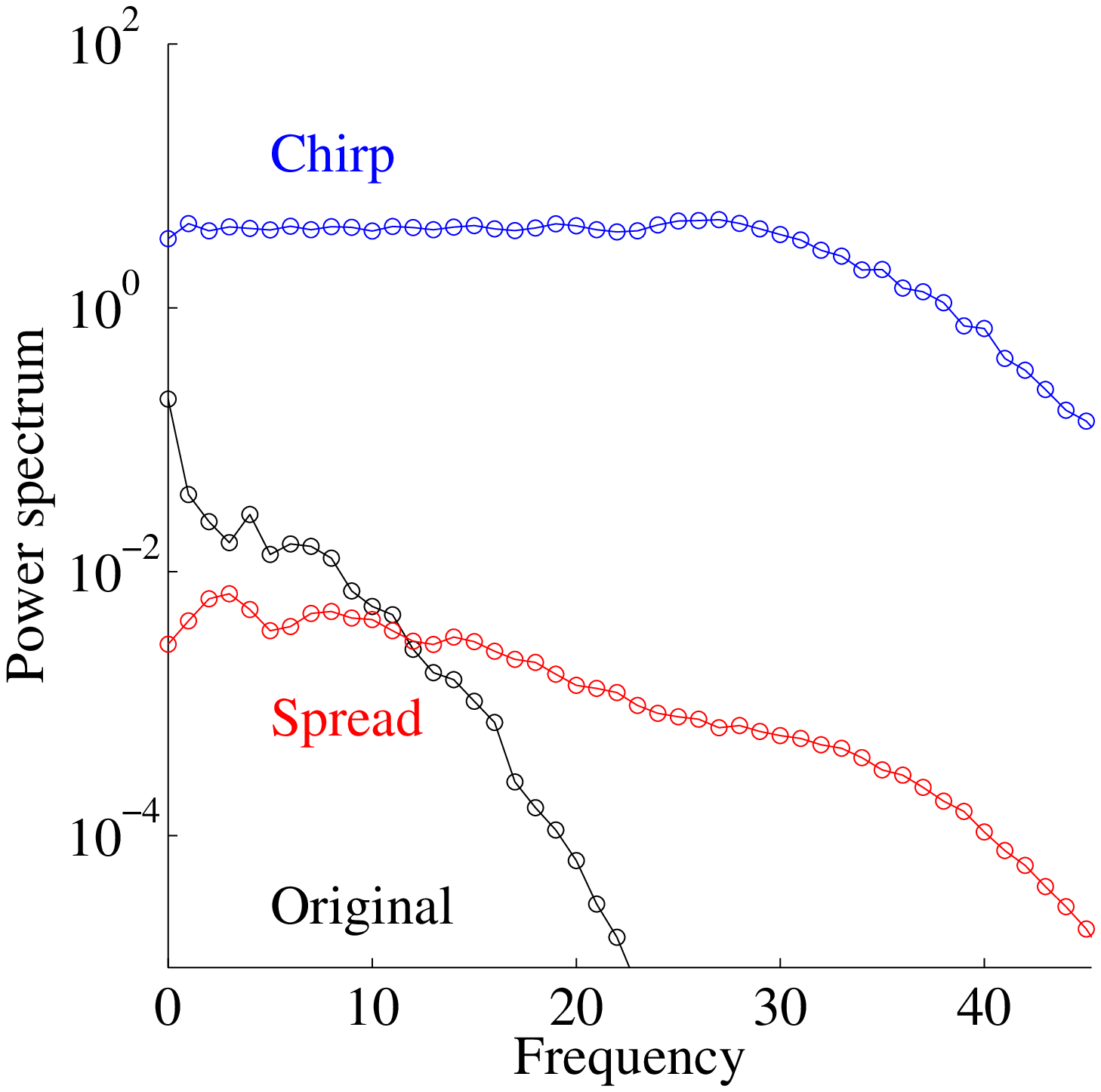}
\hspace{0.2cm}
\psfrag{Frequency}{ \hspace{-0.4cm}{\scriptsize Spatial frequency}}
\psfrag{Power spectrum}{\hspace{-0.3cm}{\scriptsize Power spectrum (${\rm iu}^2$)}}
\psfrag{Chirp}{ \hspace{0cm}{\scriptsize \color{blue}Chirp}}
\psfrag{Spread}{\hspace{2cm}{\scriptsize \color{red}Spread}}
\psfrag{Original}{\hspace{0.7cm}{\scriptsize \color{black}Original}}
\includegraphics[width=4cm,keepaspectratio]{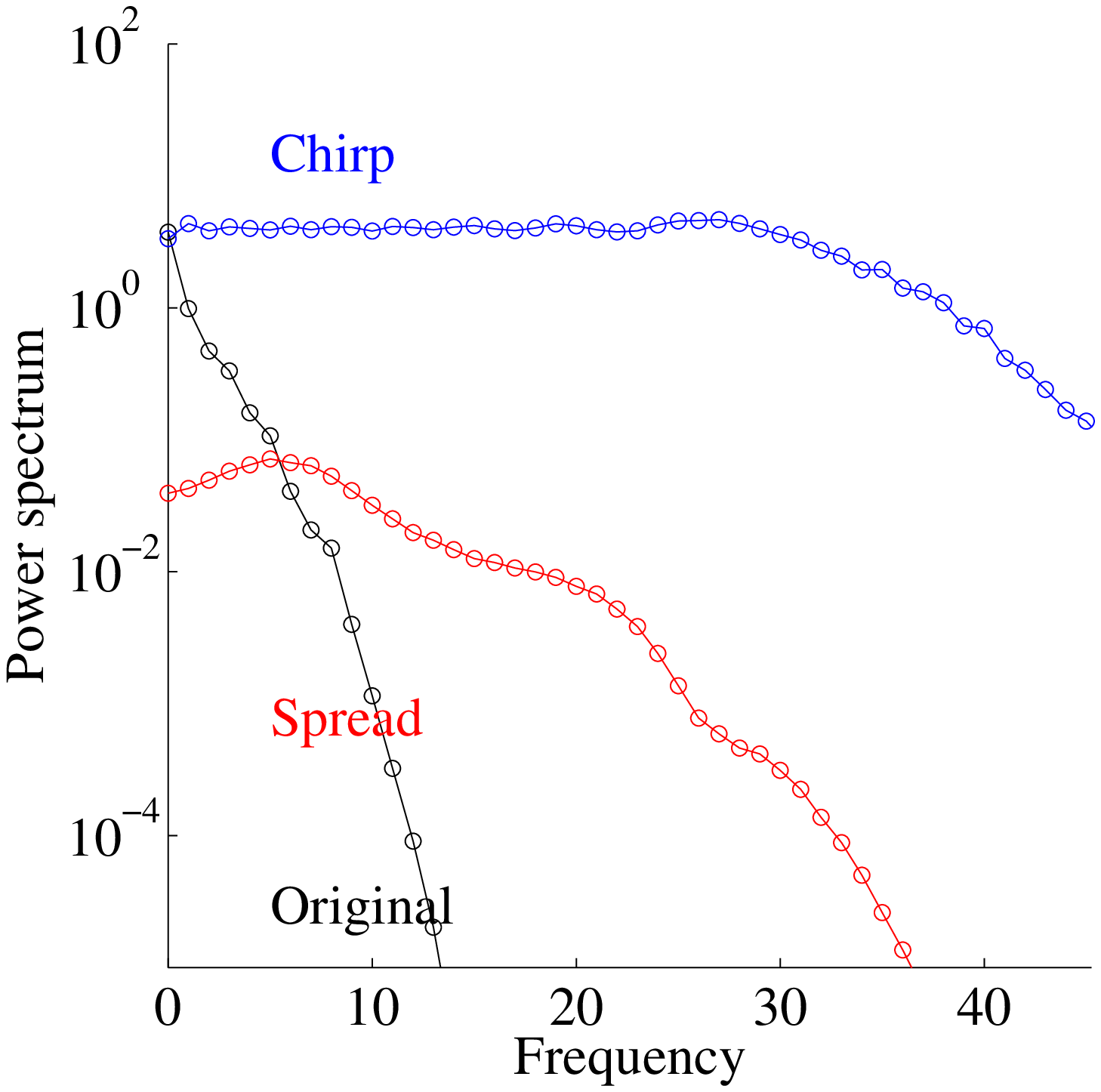}
\hspace{0.2cm}
\psfrag{Frequency}{ \hspace{-0.4cm}{\scriptsize Spatial frequency}}
\psfrag{Power spectrum}{\hspace{-0.3cm}{\scriptsize Power spectrum (${\rm iu}^2$)}}
\psfrag{Chirp}{ \hspace{0cm}{\scriptsize \color{blue}Chirp}}
\psfrag{Spread}{\hspace{2cm}{\scriptsize \color{red}Spread}}
\psfrag{Original}{\hspace{0.3cm}{\scriptsize \color{black}Original}}
\includegraphics[width=4cm,keepaspectratio]{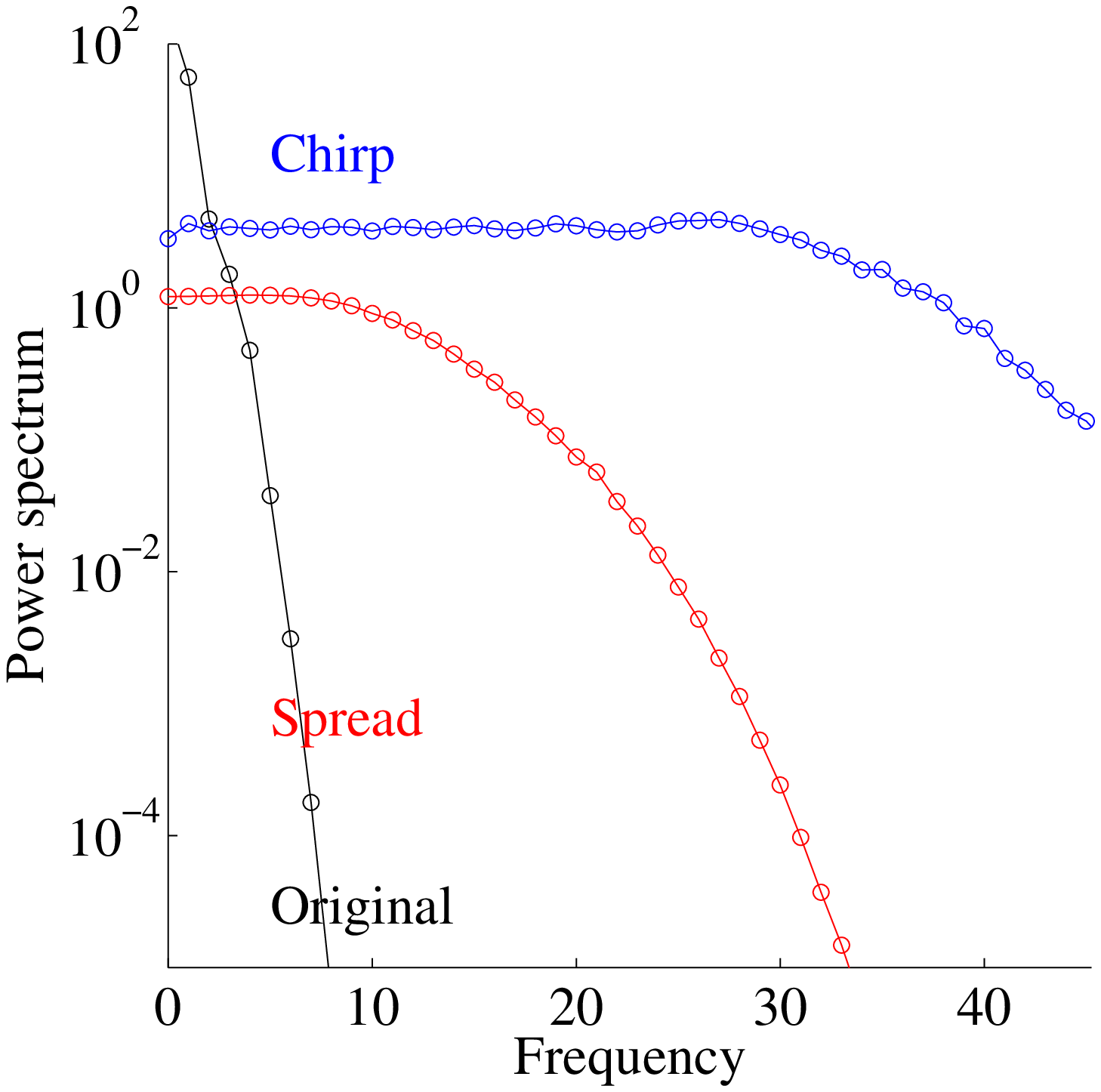}\\
\vspace{0.2cm}
\psfrag{Coverage}{ \hspace{-0.2cm}{\scriptsize Coverage}}
\psfrag{SNR (dB)}{\hspace{0cm}{\scriptsize ${\rm SNR}$ (${\rm dB}$)}}
\psfrag{G0}{ \hspace{-0.7cm}{\scriptsize $\mathsf{\Gamma}{\rm BP}_{\epsilon}0$}}
\psfrag{G1}{\hspace{-1.5cm}{\scriptsize \color{red}$\mathsf{\Gamma}{\rm BP}_{\epsilon}1$}}
\psfrag{D0}[b]{\hspace{1.3cm}{\scriptsize $\mathsf{\Delta}{\rm BP}_{\epsilon}0$}}
\psfrag{D1}[c]{\hspace{2cm}{\scriptsize \color{red}$\mathsf{\Delta}{\rm BP}_{\epsilon}1$}}
\includegraphics[width=4cm,keepaspectratio]{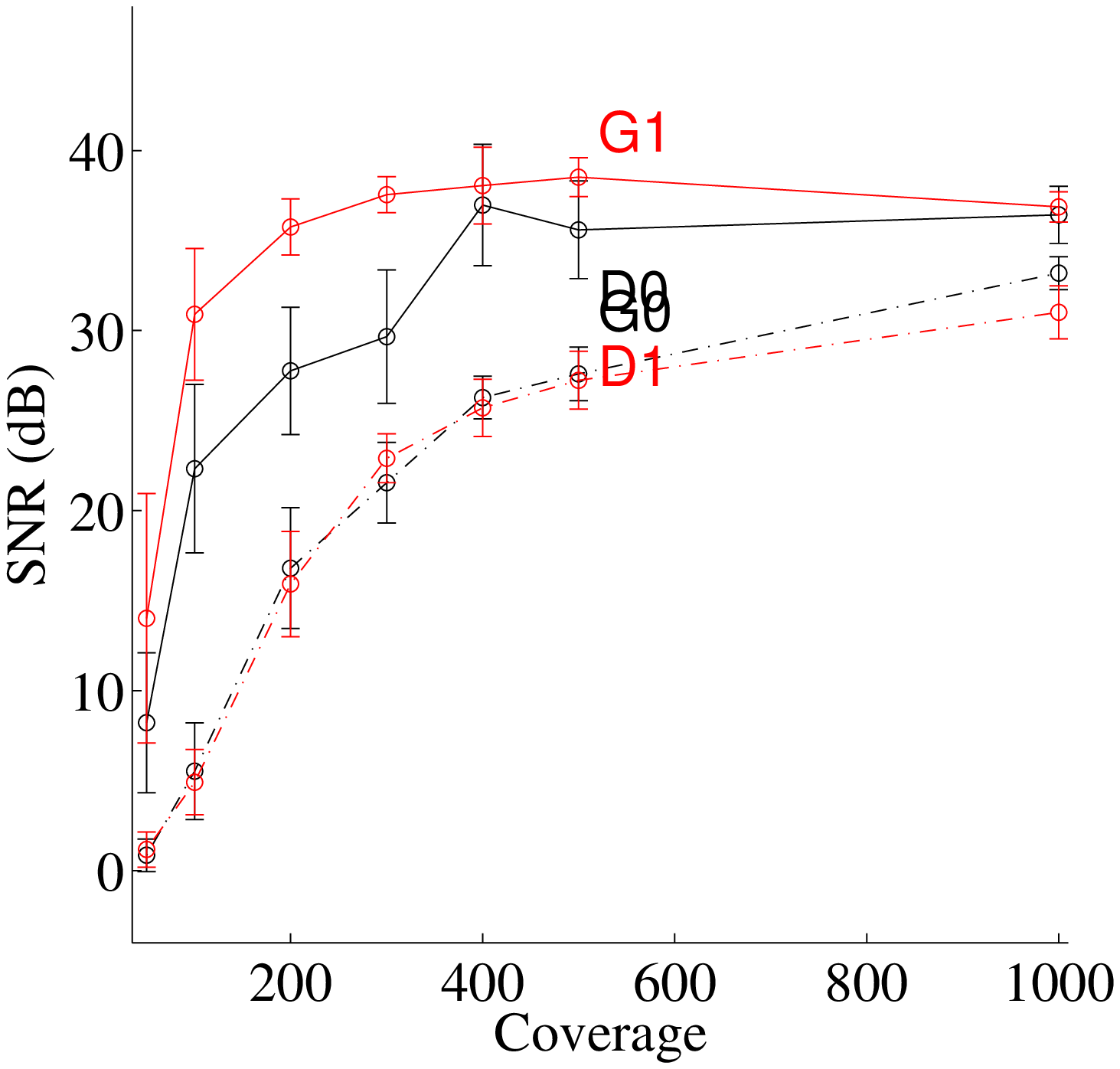}
\hspace{0.2cm}
\psfrag{Coverage}{ \hspace{-0.2cm}{\scriptsize Coverage}}
\psfrag{SNR (dB)}{\hspace{0cm}{\scriptsize ${\rm SNR}$ (${\rm dB}$)}}
\psfrag{G0}{ \hspace{0cm}{\scriptsize $\mathsf{\Gamma}{\rm BP}_{\epsilon}0$}}
\psfrag{G1}{\hspace{-1.5cm}{\scriptsize \color{red}$\mathsf{\Gamma}{\rm BP}_{\epsilon}1$}}
\psfrag{D0}[b]{\hspace{0.2cm}{\scriptsize $\mathsf{\Delta}{\rm BP}_{\epsilon}0$}}
\psfrag{D1}[b]{\hspace{1cm}{\scriptsize \color{red}$\mathsf{\Delta}{\rm BP}_{\epsilon}1$}}
\includegraphics[width=4cm,keepaspectratio]{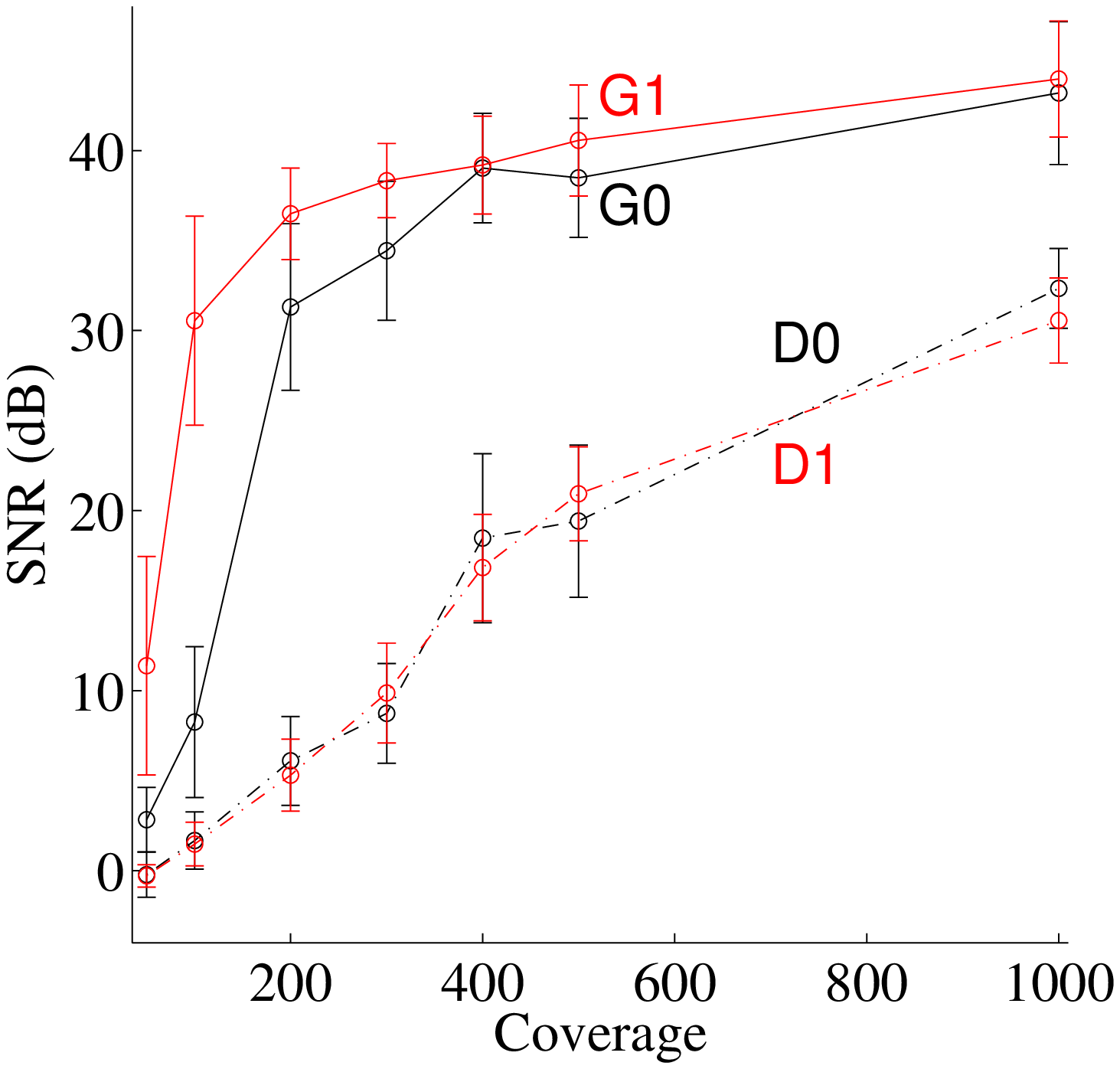}
\hspace{0.2cm}
\psfrag{Coverage}{ \hspace{-0.2cm}{\scriptsize Coverage}}
\psfrag{SNR (dB)}{\hspace{0cm}{\scriptsize ${\rm SNR}$ (${\rm dB}$)}}
\psfrag{G0}{ \hspace{0cm}{\scriptsize $\mathsf{\Gamma}{\rm BP}_{\epsilon}0$}}
\psfrag{G1}{\hspace{-1.5cm}{\scriptsize \color{red}$\mathsf{\Gamma}{\rm BP}_{\epsilon}1$}}
\psfrag{D0}[c]{\hspace{-0.5cm}{\scriptsize $\mathsf{\Delta}{\rm BP}_{\epsilon}0$}}
\psfrag{D1}[t]{\hspace{-0.1cm}{\scriptsize \color{red}$\mathsf{\Delta}{\rm BP}_{\epsilon}1$}}
\includegraphics[width=4cm,keepaspectratio]{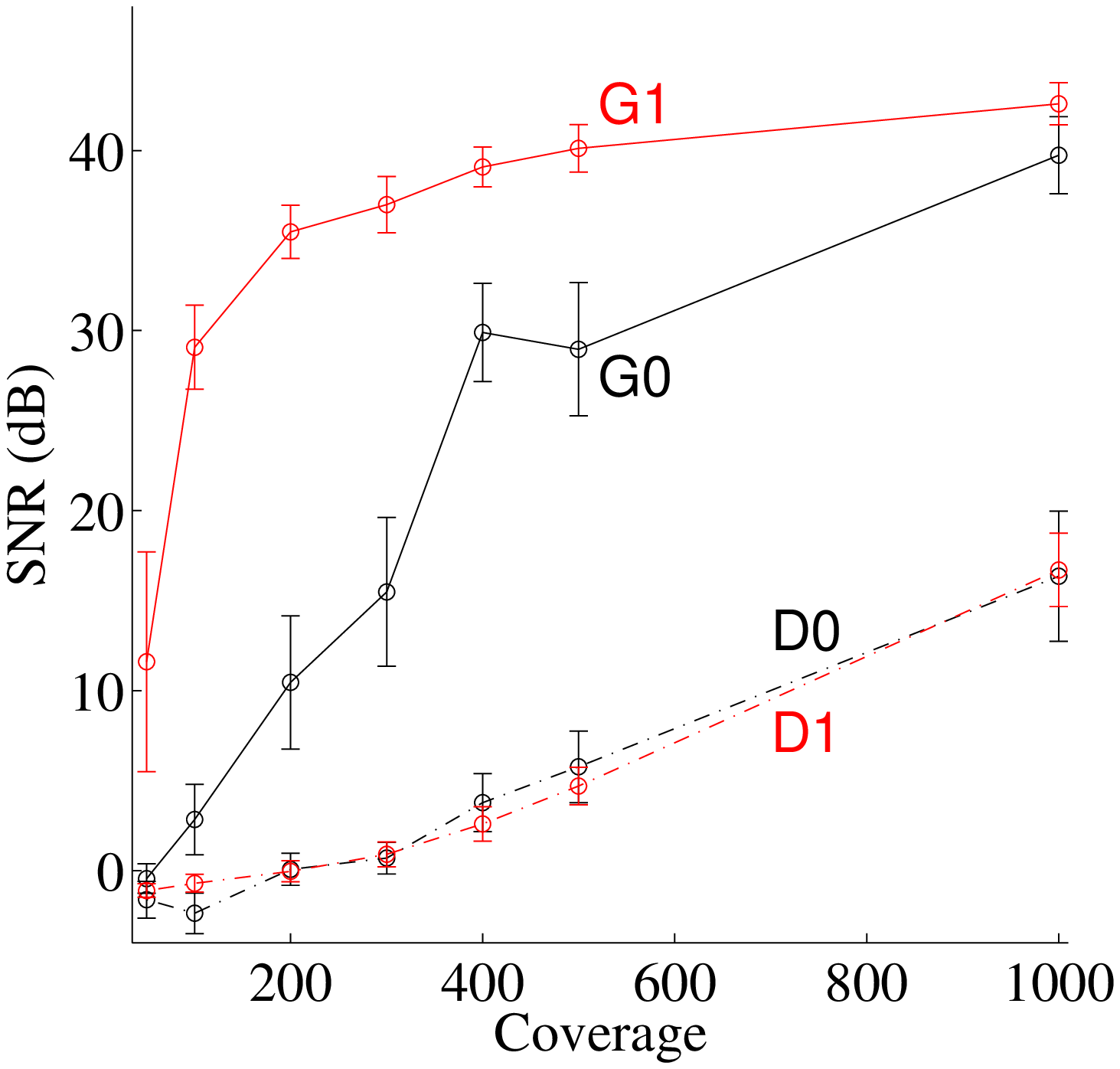}
\hspace{0.2cm}
\psfrag{Coverage}{ \hspace{-0.2cm}{\scriptsize Coverage}}
\psfrag{SNR (dB)}{\hspace{0cm}{\scriptsize ${\rm SNR}$ (${\rm dB}$)}}
\psfrag{G0}{ \hspace{0cm}{\scriptsize $\mathsf{\Gamma}{\rm BP}_{\epsilon}0$}}
\psfrag{G1}{\hspace{-1.5cm}{\scriptsize \color{red}$\mathsf{\Gamma}{\rm BP}_{\epsilon}1$}}
\psfrag{D0}{\hspace{-0.8cm}{\scriptsize $\mathsf{\Delta}{\rm BP}_{\epsilon}0$}}
\psfrag{D1}[c]{\hspace{0.5cm}{\scriptsize \color{red}$\mathsf{\Delta}{\rm BP}_{\epsilon}1$}}
\includegraphics[width=4cm,keepaspectratio]{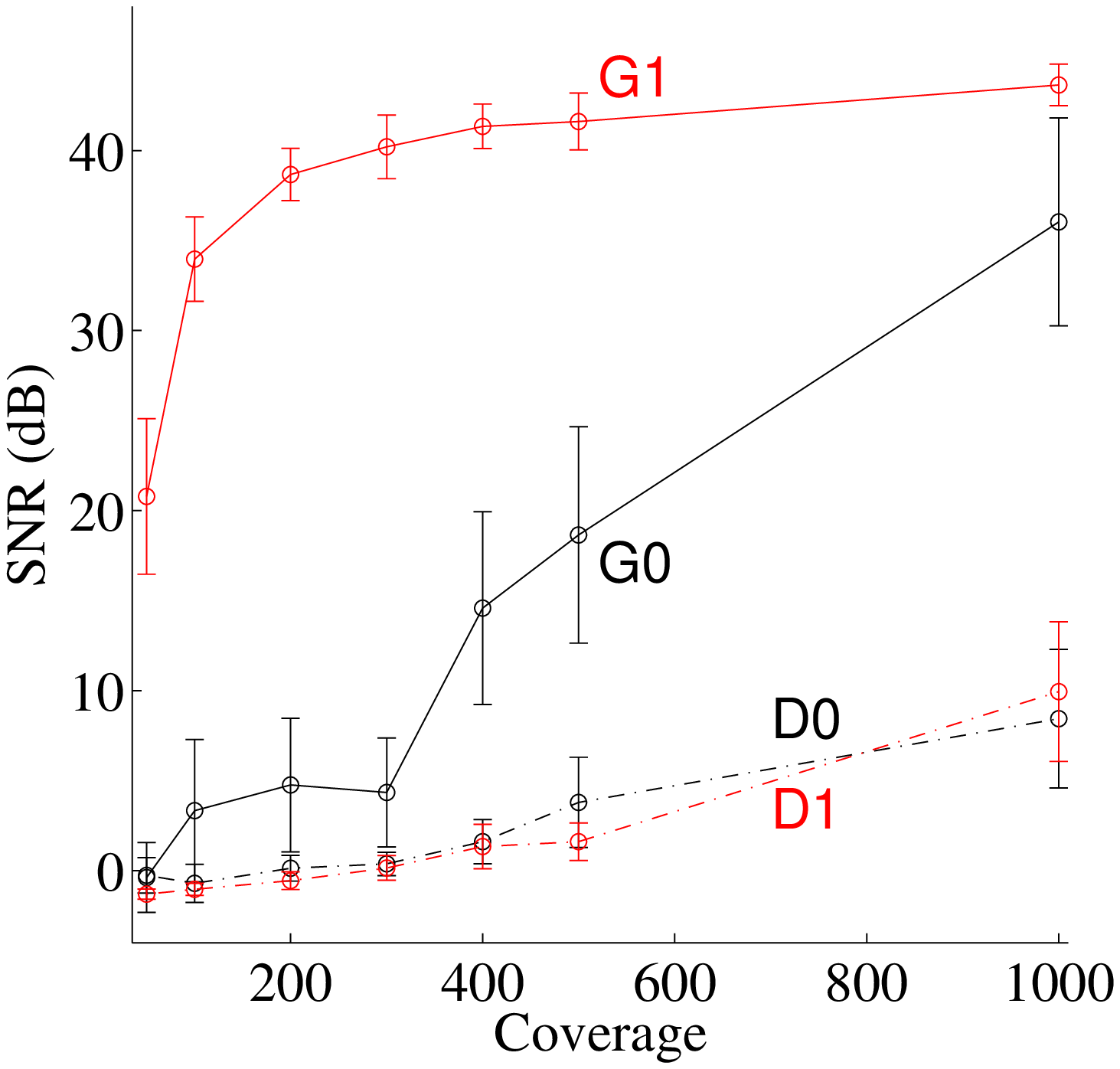}
\end{center}
\caption{\label{fig:all}From left to right, maps and graphs are associated with increasing values of the size $t_d \in \{2,4,8,16\}$ of the Gaussian waveforms constituting the sparsity dictionaries $\mathsf{\Gamma}^{(t)}$ in which the astrophysical signals considered are defined. The top panels represent original signal samples multiplied by the primary beam. Dark and light regions respectively correspond to large and small intensity values. The middle panels represent the one-dimensional sample power spectra in the square of the intensity units (${\rm iu}^2$) and as a function of the radial frequency $L^{1/2}\vert\bm{u}\vert$ with $0 \leq L^{1/2}\vert \bm{u}\vert\leq32\sqrt{2}$ for the original signals multiplied by the primary beam in the above panel (continuous black curve), for the chirp modulation (continuous blue curve), and for the signals spread by the chirp modulation (continuous red curve). The bottom panels represent the ${\rm SNR}$ of the reconstructions multiplied by the primary beam as a function of the coverage identified by the number of complex visibilities $M/2\in\{50,100,200,300,400,500,1000\}$. The curves correspond to the $\mathsf{\Delta}{\rm BP}_{\epsilon}0$ reconstructions (dot-dashed black curve), the $\mathsf{\Delta}{\rm BP}_{\epsilon}1$ reconstructions (dot-dashed red curve), the $\mathsf{\Gamma}{\rm BP}_{\epsilon}0$ reconstructions (continuous black curve), and the $\mathsf{\Gamma}{\rm BP}_{\epsilon}1$ reconstructions (continuous red curve). The latter illustrate the spread spectrum universality relative to the sparsity dictionary. Each curve precisely represents the mean ${\rm SNR}$ over the $30$ simulations, and the vertical lines identify the error at $1$ standard deviation.}
\end{figure*}
\begin{figure}
\begin{center}
\includegraphics[width=4cm,keepaspectratio]{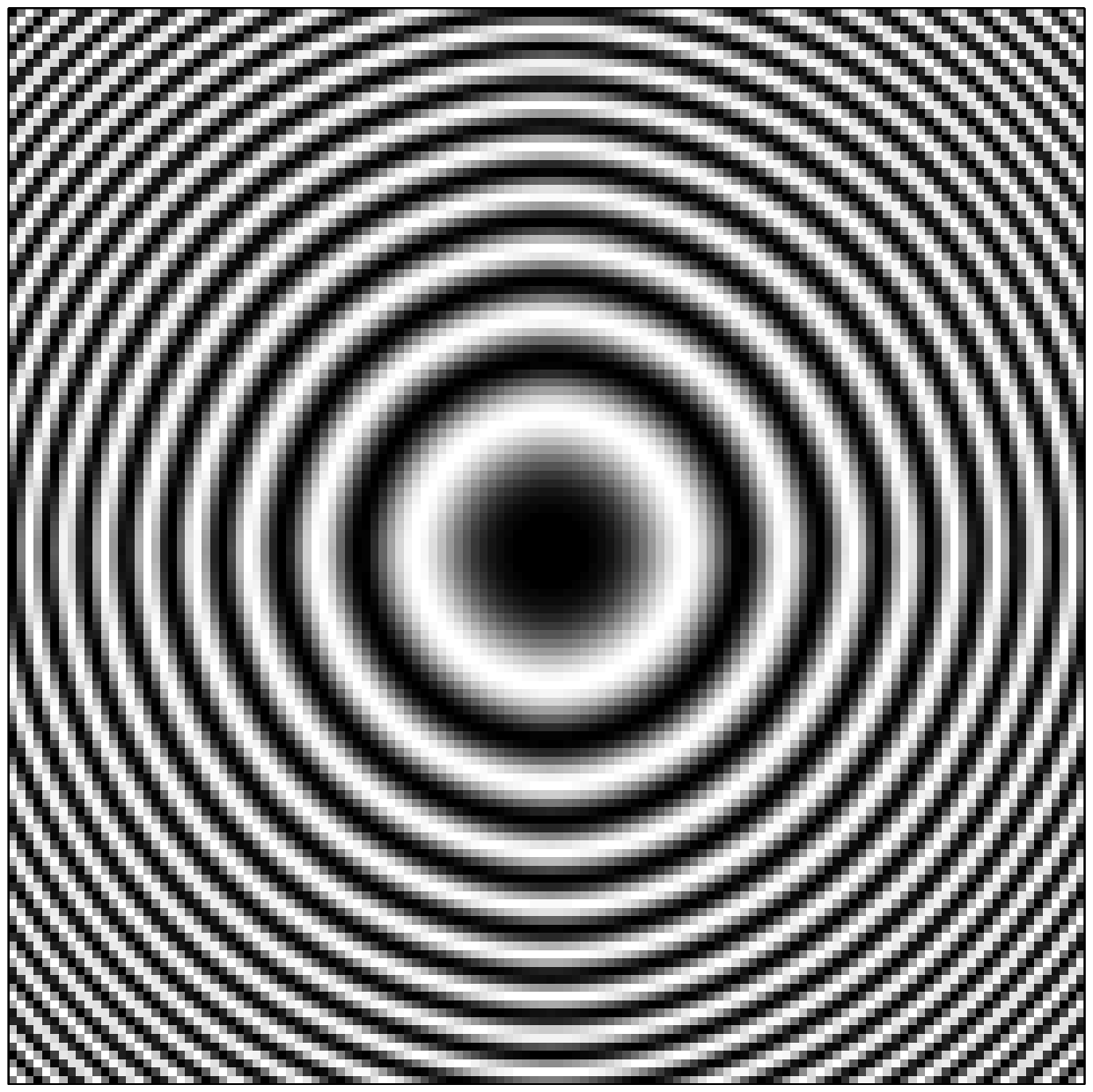}
\hspace{0.2cm}
\includegraphics[width=4cm,keepaspectratio]{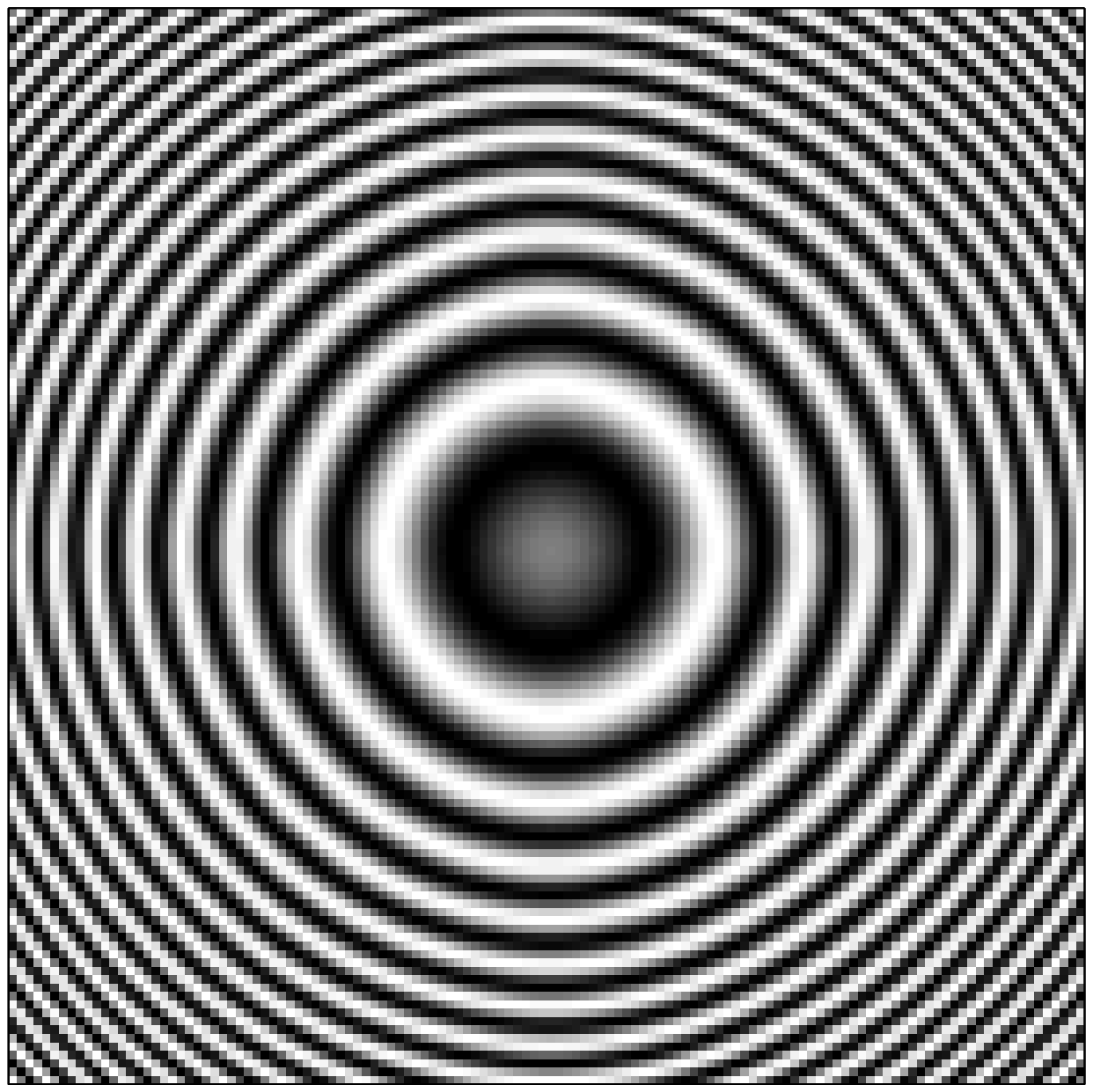}
\end{center}
\caption{\label{fig:chirp}Real part (left panel) and imaginary part (right panel) of the chirp modulation with $w_d=1$ assumed in the simulations. The chirp is sampled on a grid of $4N$ pixels in order to avoid any aliasing artifact. Dark and light regions respectively correspond to positive and negative values.}
\end{figure}
\begin{table}
\begin{center}\begin{tabular}{ccccc}
\hline 
\noalign{\vskip\doublerulesep}
Gaussian size & $t_d=2$  & $t_d=4$  & $t_d=8$  & $t_d=16$  \\
\hline
Coherence for $ w_d=0$ & $0.0518$ & $0.103$ & $0.205$  & $0.400$\\ 
Coherence for $ w_d=1$ & $0.0517$ & $0.102$ & $0.174$  & $0.151$\\
\hline
\end{tabular}\end{center}
\caption{\label{tab:mu} Values with three significant figures of the mutual coherence $\mu(\mathsf{FC}^{(w)}\mathsf{A}^{(t_0)},\mathsf{\Gamma}^{(t)})$ between the sparsity and sensing dictionaries, for sizes of the Gaussian waveforms identified by $t_d \in \{2,4,8,16\}$, both in the absence of chirp modulation, i.e. for $w_d=0$, and for a chirp modulation with $w_d=1$.}
\end{table}

The signals $x$ considered are built as the superposition of $K=10$ Gaussian waveforms, with random positions, and random central value in the interval $[0,1]$ in some arbitrary intensity units (${\rm iu}$). The signals are sampled as images $\bm{x}$ on a grid of $N=64\times64=4096$ pixels. Fields of view $L=L^{1/2}\times L^{1/2}$ with $L^{1/2}$ not larger than several degrees of angular opening may be  considered for the approximation of a planar signal to be valid. A primary beam $A^{(t_0)}$ with a full width at half maximum (FWHM) $2t_0(2\ln2)^{1/2}=L^{1/2}$ is also considered, centered on the signal and with unit value at the center. In the discrete setting defined in Section \ref{sub:Interferometric inverse problem}, the inverse problem is transparent to the precise value of the field of view so that we do not need to fix it. Observations are considered for $M/2$ complex visibilities with $M/2\in\{50,100,200,300,400,500,1000\}$. The corresponding coverages of the spatial frequencies $\bm{u}_b$, relative to the total number of spatial frequencies $\bm{u}_i$ in the Fourier plane up to the accessible band limit on the grids $B^{(N,L)}=N^{1/2}/2L^{1/2}$, are roughly between $1$\% and $25$\%. Instrumental noise is also added as independent identically distributed Gaussian noise. The corresponding standard deviation $\sigma^{(n_r)}$, identical for all $r$ with $1\leq r \leq M$, is set to $(\sigma^{(n_r)})^2=10^{-3}(\Sigma^{(\mathsf{FA}^{(t_0)}\bm{x})})^2$, where $\Sigma^{(\mathsf{FA}^{(t_0)}\bm{x})}$ stands for the sample standard deviation of the real and imaginary parts of the Fourier transform of the original signal multiplied by the primary beam $\mathsf{FA}^{(t_0)}\bm{x}$. Note that this noise measure is independent of the chirp modulation, which preserves the norm of the signal.

The size $t$ of the Gaussian waveforms may be written in terms of a discrete size $t_d$ as $t=t_d L^{1/2}/\pi N^{1/2}$. We consider the values $t_d \in \{2,4,8,16\}$ such that the Gaussian waveforms are well sampled on the grid. For $t_d=2$ the approximate band limit $B^{(A^{(t_0)}x)}$ of the signal is around the same value as $B^{(N,L)}$, while for $t_d=16$ it is much smaller. All values also satisfy the constraint that the size of the Gaussian waveforms remains smaller than that of the field of view: $t<t_0$. Original signal samples multiplied by the primary beam are reported in the top panels of Fig. \ref{fig:all}, one for each value of $t_d$ considered. The middle panels notably represent, for each value of $t_d$, the one-dimensional sample power spectrum of the original signal multiplied by the primary beam, i.e. the average of the two-dimensional sample power spectrum on all spatial frequencies $\bm{u}$ on the same annulus at fixed radial frequency $\vert\bm{u}\vert\equiv(u^2+v^2)^{1/2}$.

The component $w$ may also be written in terms of a discrete component $w_d$ as $w=w_d N^{1/2}/L$. We consider two extreme values. The case of baselines with negligible component $w$ is identified by $w_d=0$. The case of baselines with non-negligible and constant component $w$ is identified by $w_d = 1$, corresponding to a linear chirp modulation with a maximum instantaneous frequency $wL^{1/2}/2=B^{(N,L)}$, i.e. with an approximate band limit $B^{(C^{(w)})}$ not much larger than $B^{(N,L)}$ (see Fig. \ref{fig:chirp}). In other words the component $w$ is a factor $2/L^{1/2}$ larger than the maximum value of the components $u$ and $v$ in the plane of the signal, so that for the small fields of view assumed the baselines are strongly aligned with the pointing direction. One can check that this value of $w$ is consistent with the constraint for relations (\ref{ri2}) and (\ref{ri3}).

Table \ref{tab:mu} contains the values of the mutual coherence $\mu(\mathsf{FC}^{(w)}\mathsf{A}^{(t_0)},\mathsf{\Gamma}^{(t)})$ for the values of $t_d$ and $w_d$ considered. These values are computed numerically from the definition (\ref{cs4}). A computation from the theoretical relation (\ref{ssu1}) provides the same results, up to a relative discrepancy of $10$\% related to the fact that in our setting the primary beam is not completely contained in the field view. These numerical values are thus completely in the line of our discussion of Section \ref{sub:Theoretical coherence}. Additionally, the one-dimensional sample power spectra of the original signals multiplied by the primary beam, as well as those of the chirp modulation and of the signals spread by the chirp modulation are represented in the middle panels of the Fig. \ref{fig:all}, directly below the corresponding original signal for each value of $t_d$. These graphs illustrate the spread spectrum phenomenon due to the chirp modulation and the reduction of the mutual coherence intimately related to it. Up to some normalization the mutual coherence is indeed essentially the maximum value of the square root of the one-dimensional sample power spectrum.

\subsection{Simulations and ${\rm BP}_{\epsilon}$ reconstruction procedures}

A number of $30$ simulations is generated for each value of $t_d$ and $w_d$ considered. The visibilities are simulated,\footnote{In practice the grids considered allow the sampling at Nyquist-Shannon rate for signals with band limits up to  $B^{(N,L)}$ in each direction. This is not the case for the original signals after the chirp modulation. In order to avoid artificial aliasing effects in the discrete Fourier transform at the level of the frequencies probed, it is essential to introduce an operator increasing the resolution of the grids, by zero-padding in the Fourier plane, between the sparsity and sensing matrices.} and the signals are reconstructed through the ${\rm BP}_{\epsilon}$ problem, which is solved by convex optimization.\footnote{The algorithm is based on the Douglas-Rachford splitting method \citep{combettes07} in the framework of proximal operator theory \citep{moreau62}. Under our constant $w$ assumption, the computation complexity is driven at each iteration by the complexity of the FFT, i.e. $\mathcal{O}(N\log N)$. For the small value of $N$ considered the reconstruction typically takes less than $1$ minute on a single standard processor. The algorithm therefore easily scales to much larger values of $N$.} The quality of reconstruction is analyzed in terms of the signal-to-noise ratio of the reconstructions multiplied by the primary beam ${\rm SNR} \equiv-20\log_{10}(\Sigma^{(\mathsf{A}^{(t_0)}(\bm{x}-\bar{\bm{x}}))} / \Sigma^{(\mathsf{A}^{(t_0)}\bm{x})})$, where $\Sigma^{(\mathsf{A}^{(t_0)}\bm{x})}$ stands for the sample standard deviation of the original signal $\mathsf{A}^{(t_0)}\bm{x}$, and $\Sigma^{(\mathsf{A}^{(t_0)}(\bm{x}-\bar{\bm{x}}))}$ for that of the discrepancy signal $\mathsf{A}^{(t_0)}(\bm{x}-\bar{\bm{x}})$.

We actually compare two different settings for the reconstructions. In the first setting, called $\mathsf{\Gamma}{\rm BP}_{\epsilon}$, we assume that the Gaussian waveform dictionary with appropriate $t$ is known, and we use it explicitly as sparsity dictionary: $\mathsf{\Psi}\equiv \mathsf{\Gamma}^{(t)}$. As a consequence, the $\mathsf{\Gamma}{\rm BP}_{\epsilon}$ problem deals with the best possible sparsity value $K=10$. The reconstructions in the absence ($w_d = 0$) and in the presence ($w_d = 1$) of the chirp modulation are respectively denoted by $\mathsf{\Gamma}{\rm BP}_{\epsilon}0$ and $\mathsf{\Gamma}{\rm BP}_{\epsilon}1$. It is the precise setting in which we just brought up the spread spectrum phenomenon and its universality on the basis of considerations relative to the mutual coherence between the sensing and sparsity dictionaries. In the second setting, called $\mathsf{\Delta} {\rm BP}_{\epsilon}$, we assume that the sparsity dictionary is the real space basis: $\mathsf{\Psi}\equiv \mathsf{\Delta}$. As a consequence, the $\mathsf{\Delta} {\rm BP}_{\epsilon}$ problem deals with the best possible coherence value $\mu(\mathsf{F},\mathsf{\Delta})=N^{-1/2}$. However, the sparsity computed in real space increases drastically with the size of the Gaussian waveforms, suggesting that the reconstruction quality should clearly decrease when the Gaussian size increases. The reconstructions in the absence ($w_d = 0$) and in the presence ($w_d = 1$) of the chirp modulation are respectively denoted by $\mathsf{\Delta}{\rm BP}_{\epsilon}0$ and $\mathsf{\Delta}{\rm BP}_{\epsilon}1$. But the mutual coherence here remains unaffected by the chirp modulation, so that this modulation should fail to enhance the reconstruction quality in this case.

This analysis structure is strongly suggested by our interest in understanding the behaviour of the standard reconstruction algorithms used in radio interferometry. On the one hand, the standard CLEAN algorithm is essentially a Matching Pursuit (${\rm MP}$) algorithm \citep{mallat93,mallat98} which works by iterative removal of the so-called dirty beam, i.e. the inverse Fourier transform of the mask, in real space \citep{hogbom74,schwarz78,thompson04}. It is also known that CLEAN provides reconstruction qualities very similar to what we just called the $\mathsf{\Delta} {\rm BP}_{\epsilon}$ approach \citep{marsh87, wiaux09}. On the other hand, multi-scale versions of CLEAN can in principle account for the fact that the signal may be sparser in some multi-scale dictionary \citep{cornwell08b}. Assuming that the signals considered have a very sparse expansion in such dictionaries, the corresponding performances would probably be very close to that of our $\mathsf{\Gamma}{\rm BP}_{\epsilon}$ approach, provided that the equivalence between ${\rm MP}$ and ${\rm BP}$ still holds in a multi-scale setting.

\subsection{Results}

The results of the analysis are reported in the bottom panels of Fig. \ref{fig:all}. The curves represent the mean ${\rm SNR}$ over the $30$ simulations, and the vertical lines identify the error at $1$ standard deviation.

Firstly, for each size $t_d$ of the Gaussian waveforms and each reconstruction procedure, the ${\rm SNR}$ obviously benefits from an increase in the number $M/2$ of visibilities measured, corresponding to an increase in the information directly available on the signal.

Secondly, in the absence of chirp it clearly appears that, for each size $t_d$ of the Gaussian waveforms and each number of visibilities measured $M/2$, the $\mathsf{\Gamma}{\rm BP}_{\epsilon}0$ reconstruction exhibits a significantly better ${\rm SNR}$ than the $\mathsf{\Delta}{\rm BP}_{\epsilon}0$ reconstruction. This suggests that for the signals considered it is better to optimize the sparsity by accounting for the proper dictionary, than to optimize the mutual coherence by formally postulating that the signal lives in real space. It also appears that, for each number of visibilities measured $M/2$, the ${\rm SNR}$ of both kinds of reconstructions significantly degrades when the size $t_d$ of the Gaussian waveforms increases. This behaviour is due to the fact that either the mutual coherence, in the case of $\mathsf{\Gamma}{\rm BP}_{\epsilon}0$, or the sparsity, in the case of $\mathsf{\Delta}{\rm BP}_{\epsilon}0$, get farther from their optimal values.

Thirdly, the ${\rm SNR}$ of the $\mathsf{\Delta}{\rm BP}_{\epsilon}0$ and $\mathsf{\Delta}{\rm BP}_{\epsilon}1$ reconstructions remain undistinguishable from one another at $1$ standard deviation, independently of the size $t_d$ of the Gaussian waveforms and of the number of visibilities measured $M/2$. This illustrates the fact that the mutual coherence is already optimal in the absence of chirp modulation, so that any chirp modulation will fail to enhance the reconstruction quality. Let us also emphasize that we have implemented the standard CLEAN algorithm for comparison with this $\mathsf{\Delta}{\rm BP}_{\epsilon}$ setting. For each value of $t_d$ and $M$ and both for $w_d=0$ and $w_d=1$, the corresponding ${\rm SNR}$ of reconstruction (not shown in Fig. \ref{fig:all}) remains, as expected, undistinguishable from both the $\mathsf{\Delta}{\rm BP}_{\epsilon}0$ and $\mathsf{\Delta}{\rm BP}_{\epsilon}1$ reconstructions at $1$ standard deviation.

Fourthly, for any number of visibilities measured $M/2$ and for large enough size $t_d$ of the Gaussian waveforms, the ${\rm SNR}$ of the $\mathsf{\Gamma}{\rm BP}_{\epsilon}1$ is significantly larger than that of the $\mathsf{\Gamma}{\rm BP}_{\epsilon}0$ reconstruction. This is the spread spectrum phenomenon related to the reduction of the mutual coherence in the presence of the chirp modulation. Moreover, the ${\rm SNR}$ of the $\mathsf{\Gamma}{\rm BP}_{\epsilon}1$ reconstruction is essentially independent of the sparsity dictionary identified by $t_d$. This supports very strongly the principle of universality of the spread spectrum phenomenon relative to the sparsity dictionary, in perfect agreement with our theoretical considerations. The reconstruction quality also appears to be more stable around the mean ${\rm SNR}$ values in the presence of the chirp modulation. 

Finally, let us note that our pure considerations on sparsity and mutual coherence led to relation (\ref{ssu3}), which suggests the spread spectrum universality and associated optimal reconstructions in the $\mathsf{\Gamma}{\rm BP}_{\epsilon}$ setting in the limit of an infinite chirp rate. In practice though, the ${\rm SNR}$ of a $\mathsf{\Gamma}{\rm BP}_{\epsilon}$ reconstruction should saturate at some finite value, and progressively degrade above this value, due to a leakage phenomenon. The larger $w_d$, the larger the spreading of the spectrum of the signal. But the extent of the Fourier coverages considered is limited by the band limit $B^{(N,L)}$. Consequently, when the band limit of the signal after chirp modulation significantly exceeds the band limit where the visibility distributions are defined, a significant part of the energy of the signal remains unprobed.

Simulations and reconstructions have actually been performed for a range of chirp modulations with chirp rates between $w_d=0$ and $w_d=1$, as well as for a higher chirp rate $w_d=1.5$, in both the  $\mathsf{\Gamma}{\rm BP}_{\epsilon}$  setting and the $\mathsf{\Delta}{\rm BP}_{\epsilon}$ setting. In the $\mathsf{\Gamma}{\rm BP}_{\epsilon}$ setting, the ${\rm SNR}$ of reconstruction (not shown in Fig. \ref{fig:all}) undergoes a natural continuous increase for values $w_d\leq1$ from the $\mathsf{\Gamma}{\rm BP}_{\epsilon}0$ curve to the $\mathsf{\Gamma}{\rm BP}_{\epsilon}1$ curve, confirming the spread spectrum phenomenon for all values of $t_d$ and $M$ considered. The saturation of the ${\rm SNR}$ occurs around $w_d=1$, and its degradation is for example already significant for a chirp rate $w_d=1.5$, for $t_d=2$ and all values of $M$ considered. In the $\mathsf{\Delta}{\rm BP}_{\epsilon}$ setting, the ${\rm SNR}$ of reconstruction (not shown in Fig. \ref{fig:all}) exhibits no evolution for values $w_d\leq1$, independently of the values of $t_d$ and $M$ considered, confirming the fact that the chirp modulation has no impact on the mutual coherence. However, the leakage phenomenon obviously also affects the reconstructions. The corresponding degradation is observed in the same conditions as for the $\mathsf{\Gamma}{\rm BP}_{\epsilon}$ setting.

\subsection{Comments}

A huge amount of work still needs to be envisaged for analyzing the effect of baselines with non-negligible component $w$ in radio interferometry. We comment here on three important points.

Firstly, all our results should be confirmed for various levels of instrumental noise before stronger conclusions are drawn. In particular, possible implications of the spreading of the signal energy on all spatial frequencies probed due to the chirp modulation should be studied as a function of the noise level on each visibility.

Secondly, we assumed perfect knowledge of a sparsity dictionary made up of simple Gaussian waveforms. A large range of multi-scale dictionaries, notably wavelet frames, may be used in which a large variety of natural signals are known to be sparse or compressible. Let us however acknowledge the fact that a non-optimal choice of a sparsity dictionary will of course have an effect on the sparsity and possibly degrade the reconstruction. In this perspective, further analyses should be performed in order to assess the suitability of specific dictionaries.

For signals with sparse or compressible gradients, a TV norm may be substituted for the $\ell_{1}$ norm of the coefficients in the sparsity dictionary, in the very definition of the minimization problem. The TV norm of a signal is simply defined as the $\ell_{1}$ norm of the magnitude of its gradient \citep{rudin92}. A theoretical result of exact reconstruction holds for such TV norm minimization problems in the case of Fourier measurements of signals with exactly sparse gradients in the absence of noise \citep{candes06a}. But no proof of stability relative to noise and compressibility exists. Such minimization is also accessible through an iterative scheme from convex optimization algorithms \citep{candes05}. Even though our signals would not primarily be thought to have very sparse gradients, the Gaussian waveforms have in practice well defined contours. Preliminary reconstruction results using this scheme actually show promising reconstruction qualities.

Thirdly, further analyses should also be performed in order to understand the effect of the chirp modulation for realistic distributions of the baseline components $(u,v,w)$, and in particular for non-constant component $w$. However, let us already note from the bottom panels of Fig. \ref{fig:all} that, for each size $t_d$ of the Gaussian waveforms, the ${\rm SNR}$ of the $\mathsf{\Gamma}{\rm BP}_{\epsilon}1$ reconstruction rises very steeply with the number of visibilities measured $M/2$, before saturating around some multiple of the sparsity $K=10$. This suggests a selection procedure in cases where the total number of visibilities measured is very large relative to the sparsity considered. If a suitable fraction of visibilities is associated with the suitable coverage of frequencies $\bm{u}_b$ in the Fourier plane and with a suitable identical value of $w$, only these visibilities might be retained in the problem. In such a case, our present considerations could apply directly.

Let us also recall that the value $w_d=1$ requires a strong alignment of the baselines with the pointing direction for small fields of view, possibly corresponding to an unrealistically large component $w$. In this respect, the $\mathsf{\Gamma}{\rm BP}_{\epsilon}1$ reconstruction results are asymptotic values. However, a given baseline component $w$ will actually correspond to a larger $w_d$ on a larger field of view $L$ and for a given band limit $B^{(N,L)}$, so that $w_d=1$ will become more realistic. This is actually the exact opposite argument to the one used for the definition of faceting algorithms, which decompose a given field of view on subfields where the effect of $w$ is negligible. In that regard, the extension of our results on wide fields of view on the celestial sphere will be essential \citep{cornwell08a,mcewen08}, notably with regard to forthcoming radio interferometers such as the Square Kilometer Array (SKA) \citep{carilli04}.\footnote{http://www.skatelescope.org/}

\section{Conclusion}
\label{sec:Conclusion}

We have focused our attention on radio interferometers with small field of view and baselines with non-negligible and constant component in the pointing direction, for which a linear chirp modulation affects the astrophysical signals probed. Considering simple sparse signals made up of Gaussian waveforms, we have discussed the sensitivity of imaging techniques relative to the sparsity dictionary in the context of the theory of compressed sensing. A theoretical computation of the mutual coherence between the sparsity and sensing dictionaries, as well as the results of our numerical simulations, suggest the universality of the spread spectrum phenomenon relative to the sparsity dictionary, in terms of the achievable quality of reconstruction through the ${\rm BP}_{\epsilon}$ problem.

\section*{Acknowledgments}

The authors wish to thank L. Jacques for productive discussions as well as M. J. Fadili for private communication of results on optimization by proximal methods. The authors also thank the reviewer for his valuable comments. Y. W. is Postdoctoral Researcher of the Belgian National Science Foundation (F.R.S.-FNRS). Y. B. a Postdoctoral Researcher funded by the APIDIS European Project.

\label{lastpage}

\end{document}